\documentclass[%
 reprint,
 twocolumn,
 superscriptaddress,
 showkeys,
 amsmath,amssymb,
 aps,
 pre,
]{revtex4-2}

\usepackage{graphicx}% Include figure files
\usepackage{dcolumn}% Align table columns on decimal point
\usepackage{bm}% bold math
\usepackage[english]{babel}
\usepackage[utf8]{inputenc}
\usepackage{hyperref} % For hyperlinks in 
\usepackage{amsthm,amsmath}
\usepackage{array,booktabs,tabularx}
\usepackage{graphicx}
\usepackage{multirow}
\usepackage{color}
\usepackage{float}
\usepackage{ragged2e,tabularx, makecell}
\usepackage{relsize}
\usepackage{txfonts}
\usepackage{verbatim}
\usepackage{cprotect}
\usepackage{etoolbox}
\usepackage[pass]{geometry}
\usepackage{setspace}
\usepackage{changepage}
\usepackage{CJK}
\usepackage{titlesec}
\usepackage{url}
\usepackage{natbib}

\definecolor{Green}{RGB}{34, 139, 34}
\definecolor{dkgreen}{rgb}{0,0.6,0}
\definecolor{orange}{rgb}{1,0.5,0}
\definecolor{darkblue}{rgb}{1,0.7,1.0}

\hypersetup{pdftoolbar=true, pdfmenubar=true, pdffitwindow=false, pdfstartview={FitH}, pdfcreator={}, pdfproducer={}, pdfnewwindow=true, colorlinks=true, linkcolor=red, citecolor=Green, filecolor=magenta, urlcolor=cyan,}

\begin{document}

\title{The Dynamics of Faculty Hiring Networks}

\author{Eun Lee}
    \email[Corresponding author: ]{eule4020@colorado.edu}   % email address
    \affiliation{Department of Computer Science, University of Colorado Boulder, Engineering Drive, 80309 Boulder, CO, USA}
    \affiliation{BioFrontiers Institute, University of Colorado Boulder, Colorado Ave., 80303 Boulder, CO, USA}
   
\author{Aaron Clauset}
    \thanks{These two authors contributed equally}
    \affiliation{Department of Computer Science, University of Colorado Boulder, Engineering Drive, 80309 Boulder, CO, USA}
    \affiliation{BioFrontiers Institute, University of Colorado Boulder, Colorado Ave., 80303 Boulder, CO, USA}
    \affiliation{Santa Fe Institute, Hyde Park Road, 87501 Santa Fe, NM, USA}

\author{Daniel B. Larremore}
    \thanks{These two authors contributed equally}
    \affiliation{Department of Computer Science, University of Colorado Boulder, Engineering Drive, 80309 Boulder, CO, USA}
    \affiliation{BioFrontiers Institute, University of Colorado Boulder, Colorado Ave., 80303 Boulder, CO, USA}

\begin{abstract}
Faculty hiring networks—who hires whose graduates as faculty—exhibit steep hierarchies, which can reinforce both social and epistemic inequalities in academia. Understanding the mechanisms driving these patterns would inform efforts to diversify the academy and shed new light on the role of hiring in shaping which scientific discoveries are made. Here, we investigate the degree to which structural mechanisms can explain hierarchy and other network characteristics observed in empirical faculty hiring networks. We study a family of adaptive rewiring network models, which reinforce institutional prestige within the hierarchy in five distinct ways. 
Each mechanism determines the probability that a new hire comes from a particular institution according to that institution's prestige score, which is inferred from the hiring network's existing structure. 
We find that structural inequalities and centrality patterns in real hiring networks are best reproduced by a mechanism of global placement power, in which a new hire is drawn from a particular institution in proportion to the number of previously drawn hires anywhere. 
On the other hand, network measures of biased visibility are better recapitulated by a mechanism of local placement power, in which a new hire is drawn from a particular institution in proportion to the number of its previous hires already present at the hiring institution.
These contrasting results suggest that the underlying structural mechanism reinforcing hierarchies in faculty hiring networks is a mixture of global and local preference for institutional prestige. 
Under these dynamics, we show that each institution's position in the hierarchy is remarkably stable, due to a dynamic competition that overwhelmingly favors more prestigious institutions. These results highlight the reinforcing effects of a prestige-based faculty hiring system, and the importance of understanding its ramifications on diversity and innovation in academia.
\end{abstract}

\keywords{Faculty hiring, Prestige hierarchy, Inequality, Network modeling, Hiring mechanism}%Use showkeys class option if keyword
                              %display desired
\maketitle

%%%%%%%%%%%%%%%%%%%%%%% Introduction %%%%%%%%%%%%%%%%%%%%%%%%%%%%
\section*{Introduction}

Faculty hiring is a crucial process that shapes the composition and structure of the academic workforce. When one department hires a graduate of another department as faculty, it represents an implicit endorsement of the doctoral institution. One aspect of a department's power is thus its ability to place its graduates as faculty at other institutions. This placement power can be inferred from analyzing the structure of faculty hiring networks, in which nodes represent departments and a directed link 
$(i \rightarrow j)$ indicates that of a graduate of node $i$ is faculty at node $j$.  
Across academic disciplines, faculty hiring networks exhibit a few highly stereotyped properties, which differentiate them from other types of directed and weighted networks. First, out-degree distributions are heavy-tailed, indicating that most faculty graduate from a relatively small number of institutions, while low-variance in-degree distributions indicate that, within each field, departments do not vary dramatically in size~\cite{clauset2015sciadvan}. Second, faculty hiring networks are well described by a steep linear hierarchy of the nodes in which 86-91\% of directed edges point from higher-ranked nodes toward lower-ranked nodes,
and the strength of this hierarchy is much greater than can be attributed to the heavy-tailed out-degree distribution alone~\cite{clauset2015sciadvan}. 

Prestige hierarchies have been shown to drive numerous inequalities.
For example, the top ten most prestigious institutions in the Humanities produce over half (51.3\%) of all published articles in top journals, leading to far greater visibility for faculty at the top~\cite{Piper2017Epistemicineq}.
Institutional prestige also drives both the productivity and impact of early-career academics~\cite{Way2019PNAS}.~These effects, in turn, strengthen the existing prestige hierarchy that produced them in the first place~\cite{clauset2015sciadvan,samway2016www}, creating a feedback loop in which network position becomes self-reinforcing.
Furthermore, because placement power is so skewed across institutions~\cite{clauset2015sciadvan}, the demographics and research specialities of highly ranked PhD programs will ultimately shape the demographic composition~\cite{samway2016www} and research agendas of entire fields~\cite{morgan2018EPJB,hofstra2020pnas}. 
Hence, a deeper understanding of the mechanisms that create and maintain prestige hierarchies would have broad ramifications for research on scientific productivity, citation patterns, the processes that drive scientific discovery, and efforts to diversify the academy along various dimensions. 

Network models have shed light on the mechanisms that create hierarchies in many social systems.~For instance, {\it a posteriori} analysis of hierarchies in high school friendship networks revealed that both unreciprocated and reciprocated relationships are critical to explaining observed network structure, yet they follow different attachment mechanisms~\cite{Ball2012Friendshipstatus}. Other models, focused on the dynamics of hierarchies over time, have incorporated nodes' preferences into the process by which new directed links are formed~\cite{Larremore_2021_PNAS}, bringing together network formation and discrete choice modeling~\cite{overgoor2019choosing,feinberg2020choices}. Importantly, these statistically generative models allow for model comparison in explaining empirical data.~In economics, models based on iterative games have been shown to capture hierarchical nested structures~\cite{Koing2014Nestedness}. These studies are distinct in that they focus on network formation in the absence of growth~\cite{Larremore_2021_PNAS, Koing2014Nestedness}, which differentiates them from the well studied models of network attachment~\cite{price1965networks,Price_model_JASS}. 

While this body of past work has provided general insights into hierarchy formation and dynamics, the particular mechanisms underlying prestige hierarchies in faculty hiring remain unclear.~In this study, we specifically focus on understanding which mechanisms can or cannot describe the hierarchical structures common to faculty hiring networks.~We explore a diverse set of hiring network models with feedback, which capture a variety of plausible dynamics in real hiring processes. In particular, we employ rewiring network models in which edges are rewired over time, representing retirement and hiring, based on specific network-based conceptualizations of institutional prestige. 

To anchor this work in the real-world properties of faculty hiring networks, we study our models in the context of empirical faculty hiring data from Business (BS), Computer Science (CS), and History (HS), previously hand collected and curated~\cite{clauset2015sciadvan}. Comparing model-generated networks with empirical networks allows us to avoid pursuing unrealistic link formation mechanisms. 
By simulating network dynamics from empirically supported mechanisms, we then investigate the  mobility of individual institutions within prestige hierarchies over time to deepen our understanding of the factors that lead to the steepening or flattening of network hierarchies more broadly.    %I believe leaving the sections in separate files is more organized, change it if you desire 

\section*{Methods}
\label{sec:methods}

%%%%%%%%%%%%%%%%
\subsection*{Empirical Hiring Networks}
Using a dataset containing the education and employment histories of more than $16,000$ tenure-track or tenured faculty at $461$ PhD-granting Business (BS), Computer Science (CS), and History (HS) departments in the U.S. and Canada~\cite{clauset2015sciadvan,systematicineqdata} (Supplementary Table~\ref{sup_tab:network_summary}), we constructed three directed hiring networks, one for each field. We represent each network as a $N \times N$ matrix $\mathbf{A}$ in which the number of edges from $i$ to $j$ corresponds to the number of faculty at institution $j$ who obtained a PhD from institution $i$. There are $N=112$ nodes (institutions) and $9,336$ faculty in the BS network, $N=205$ nodes and $5,032$ faculty in the CS network, and $N=144$ nodes and $4,556$ faculty in the HS network. 

%%%%%%%%%%%%%%%%%%%%%%%%%%%%%%%

\begin{figure}
\begin{tabular}{l}
\\
\includegraphics[width=0.95\linewidth]{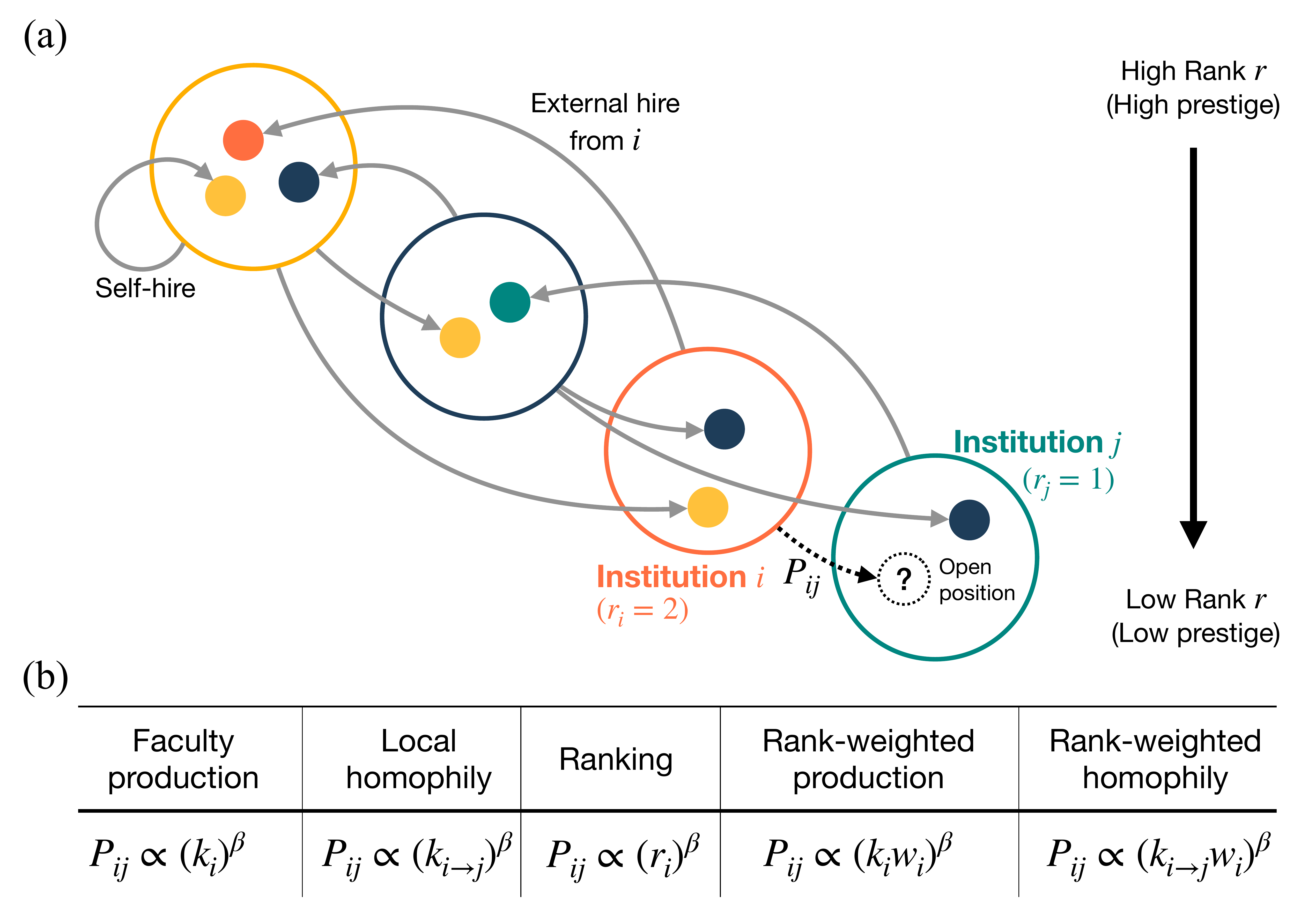}\\
\end{tabular}
\caption{{\textbf{Schematic representation of five simple models for faculty hiring.}} (a) Four large circles represent institutions, and filled circles represent faculty who graduated from the color-matched institutions. Grey links denote faculty hiring, including self-hiring. When a department $j$ has an open position (a retirement), a new hiring event from institution $i$ will occur with probability $P_{ij}$, determined by a hiring model. All institutions are assigned a prestige rank, where a high rank $r_i$ denotes high prestige, calculated via SpringRank~\cite{DeBacco2018Sciadv}. (b) Mathematical definition of $P_{ij}$ for five hiring models, showing how each model formalizes a different notion of institutional prestige: the number of faculty produced by an institution $i$ ($k_i$), the number of previous hires from $i$ at $j$ ($k_{i\rightarrow j}$), the rank of $i$ ($r_{i}$), the weighted number of faculty produced by $i$ ($k_{i}w_i$), and the weighted number of previous hires from $i$ at $j$ ($k_{i\rightarrow  j}w_i$) with a preference strength $\beta$ of an institution's prestige.}
\label{fig:schematic}
\end{figure}

\subsection*{Models of Faculty Hiring}
To model the dynamics of faculty hiring networks, we considered an evolving adaptive rewiring model that represents the processes of retirement and hiring, in which the network's current structure endogenously affects future hiring decisions. In each time step, a randomly selected edge $(x \rightarrow j)$ is removed, which signifies a retirement that creates an open position at a department $j$. Then, the open position at $j$ is filled by choosing a new hire from a node $i$ with probability $P_{ij}$. All institutions in the network follow the same hiring model for new hires. Values of $P_{ij}$ are determined according to one of five hiring mechanisms (Fig.~\ref{fig:schematic}(b)):

\begin{enumerate}
\item Faculty production:~An institution $j$ prefers an institution $i$ based on the total number of faculty $k_i$ produced by $i$. As the number of faculty alumni of an institution anywhere in the network increases, the probability of that institution placing a new graduate as faculty grows proportionally [$P_{ij} \propto (k_i)^{\beta}$].
\item Local homophily:~An institution $j$ prefers an institution $i$ in proportion to the number of faculty at $j$ that are already from $i$ ($k_{i\rightarrow j}$). This localized preference represents a historical preference to hire again from the specific institutions that have placed graduates at $j$ [$P_{ij}\propto (k_{i\rightarrow j})^{\beta}$]. 
\item Ranking:~An institution $j$ prefers an institution $i$ in proportion to $i$'s rank $r_i$. This mechanism assumes a well-known ranking $r_i$ of each institution in a field, but which may be updated over time [$P_{ij}\propto (r_i)^{\beta}$].
\item Rank-weighted production:~An institution $j$ considers institution $i$'s placement power (the number of faculty from the institution $i$) and its ranking together, such that the probability is proportional to a rank-weighted number of $k_i$ faculty graduates from $i$, where $w_i = r_i^\alpha/ \sum_{v\in{N}} r_v^\alpha$, here we set $\alpha =1$ for simplicity [$P_{ij}\propto (k_i w_i)^{\beta}$]. 
\item Rank-weighted homophily:~An institution $j$ considers a combination  of the number of faculty at $j$ that are specifically from $i$ and $i$'s rank weight $w_i$ [$P_{ij}\propto (k_{i\rightarrow j} w_i)^{\beta}$].
\end{enumerate}

Across all five models, we use a ``strength" parameter $\beta \geq 0$ to control the salience of the particular definition of institutional prestige. When $\beta = 0$, $P_{ij} = 1/N$ for all institutions, and thus, each new hire will be a uniformly random choice. As a result, this point in parameter space is common to all five models. When $\beta = 1$, hiring is proportional to the institutional prestige as defined. As $\beta$ increases, the probability becomes more concentrated on the most prestigious institutions, and in the limit of $\beta \rightarrow \infty$, every new hire is made from only the single most prestigious institution. 

We also use a parameter $p$ to control the frequency with which a particular prestige-based hiring mechanism is applied. At each time step after an edge is removed uniformly at random, the institution $j$ makes its next hire according to the prestige mechanism with probability $1-p$, and otherwise it hires by choosing $i$ uniformly random. In this way, the parameter $p$ interpolates between a fully prestige-based hiring dynamics ($p=0$) and a fully random hiring dynamics ($p=1$). As with $\beta = 0$, the point $p=1$ in parameter space produces the same, uniformly random hiring dynamics. 

Access all five models, a parameterization for $\beta$ and $p$ fully specifies a hiring dynamics model, and we are interested in understanding which points or regions in this space produce realistic faculty hiring networks.

\subsection*{Inferring Hierarchies and Measuring Their Steepness}
A key evaluation of the five faculty hiring models will be the degree to which each produces realistic prestige hierarchies.~Because each model defines an evolving network, whose edges rewire over time, a node’s position or rank $r_i$ in the hierarchy may be dynamic. We track the evolution of the hierarchy by periodically estimating its structure from the current network, using the computationally efficient SpringRank method~\cite{DeBacco2018Sciadv}. This method is known to infer hierarchies that correlate well both with authoritative academic rankings and with those of other standard methods, such as the Minimum Violation Ranking (MVR) algorithm of Ref.~\cite{clauset2015sciadvan}.

For a particular hierarchy $\{r_i\}$, we define its steepness $\rho$ to be the fraction of edges that ``violate” the hierarchy.~A directed edge $j$ to $i$ is violating if it points ``up" the hierarchy, i.e., if $r_i > r_j$. Thus, the steepness $\rho$ is a ratio of the number of upward-directed edges (faculty who are hired at departments more prestigious than where they graduated) to the total number of edges. A value of $\rho = 0.5$ would indicate that the probability of a hiring edge pointing ``up" the hierarchy equals the probability of it pointing down, and the larger a value of $\rho$, the closer the network is to a perfect hierarchy, in which every edge points ``down" the hierarchy. 

In addition to the hierarchy’s steepness, we also use the Gini coefficient $G$ of the out-degree distribution as a second measure of a hierarchy’s structure. The Gini coefficient is calculated in the standard way, and directly quantifies how unequal the placement powers (out-degrees) are across institutions. We note that $G$ and $\rho$ capture different aspects of the hierarchy’s inequality. A network can have a low value of $G$ but a high value of $\rho$, and vice versa.

%%%%%%%%%%%%%%%%%%%%%%%%%%%%%%%

\subsection*{Hiring Simulations}
\label{subsec:intialization}
We systematically study the behavior of these models via simulation. In this exploration, we consider two types of initial conditions: an idealized setting in which institutions begin on equal footing with each other, and no department has advantage in terms of network position over any other; and a more realistic setting, where institutions begin with faculty alumni sets a deterministic equal in size to those we observe empirically. In both settings, we set $N$ equal to the number of institutions, and we set the institution sizes (in-degree) equal to those observed in the particular empirical hiring network. In the first setting, the PhD institution of each hire is chosen uniformly at random from among all institutions, creating a network with a Poisson out-degree distribution. 
In the second setting, the PhD institutions of faculty are chosen by selecting a uniformly random matching of in- and out-degrees, corresponding to a directed version of the configuration model, which replicates the heavy-tailed structure of the empirical out-degree distribution. 
We refer to these two initialization settings as \textit{egalitarian} and \textit{skewed}, respectively.

The process of retirement and hiring is continued until the network reaches a dynamic equilibrium, at which point the large-scale statistical structure of the network reaches a steady state even as additional edges continue to be rewired. We define reaching a steady state by the fraction of violations in the hierarchy, given by $\rho$. 
In most settings, running the simulation for $400N^2$ steps is sufficient to produce a stable value of $\rho$. However, when $p$ is small, e.g., $p\leq 0.1$, the time required to produce a similarly stable value grows in a slightly non-linear way; in these instances, we find that $800N^2$ steps is sufficient.

\subsection*{Finding $\beta$ and $p$ to Reproduce Structural Inequality}
\label{subsec:selectbetap}

To find parameter combinations of $\beta$ and $p$ that reproduce the empirically observed values of $\rho$ and $G$, we  measured the Euclidean distance between simulated and empirical networks' as, 
\begin{equation}
\label{eq:euclidean2}
  D = [G(p,\beta) - G_{\text{data}}]^2 + [\rho(p, \beta)-\rho_{\text{data}}]^2.  
\end{equation}

\noindent For each of the five models and two initial conditions, we measured the distance $D$ over a grid of $\beta$ and $p$ values. Due to the computational cost of the simulations, we found approximate minimizers of $D$ using an efficient heuristic.~Values of $\beta$ for which $D>0.1$ for all $p$ were ruled out as implausible. For each remaining value of $\beta$, we fit a function of the form $\rho(p) = ae^{-bp}+c$, and selected the value of $p$ which minimized $D$. This procedure resulted in values of $p$ and $\beta$ that best reproduced empirically observed network properties in an efficient manner (Table~\ref{Stab:bestp}).

\subsection*{Quantifying Biased Visibility}
\label{subsec:visibility}

In networks with community structure or strong linear hierarchies, nodes located in different parts of the network may have markedly different local neighborhoods. As a result, local estimates of a network's global composition may be inaccurate. 
Thus, individual actors, like the faculty at a particular institution, may form inaccurate beliefs about the composition of their field, 
because which parts of the network a node ``sees" depends on how centrally located that node is and how broadly distributed its neighbors are~\cite{mirta2012socialsampling,lerman2016Plosone}. In faculty hiring networks, this phenomenon called biased visibility, correlates with prestige ranking and produces to two related phenomena. First, because placement rates are highly skewed, favoring the most prestigious institutions, the top-ranked 10\% of institutions are likely to be far overrepresented in the local neighborhoods of all other institutions. Second, the strength of this overrepresentation varies by the rank of the perceiving institution.

We quantify the visibility bias of a set of nodes $X$ in the perception of a node $i$ by defining the quantity
\begin{equation}
    B_{i, X} = \frac{V_{i,X}-|X|/N}{|X|/N} = \frac{\sum_{j\in X} A_{ij}/k_{i}^{in} - |X|/N}{|X|/N},
    \label{eq:biasedvisibility}
\end{equation}

\noindent to be the relative representation of a set of institutions $X$ in the neighborhood of institution $i$, where $A_{ij}$ denotes the number of faculty hired at institution $i$ who graduated from $j$, and $k_{i}^{in}$ represents the number of incoming degree of node $i$.~Here, we consider three choices of the set $X$: (i) the top-ranked 10\% of institutions, (ii) the bottom-ranked 10\% of institutions, or (iii) and the set of institutions in the rank decile spanning the 50th to 60th percentiles. These three definitions of $X$ allow us to quantify the bias visibility of the most and the least prestigious institutions along with the middle of the hierarchy, providing some sense of how the whole hierarchy behaves. Positive values of $B_{i,X}>0$ indicate that $X$ is overrepresented in the neighborhood of $i$, while negative values indicate that $X$ is underrepresented in the neighborhood of $i$.

\section*{Results}
\subsection*{Reproduction of Empirical Structural Inequalities}
To compare the five hiring models' abilities to reproduce the observed values of the two measures of structural inequality, the Gini coefficient $G$ and the hierarchy steepness $\rho$, we systematically varied the strength of prestige preference $\beta \in [0.0, 2.0]$ and the level of random hiring $p \in [0.0, 1.0]$ for both the egalitarian and skewed initial conditions (see Methods). Analyzing the output of simulations leads to two key insights.

%%%%%%%%%%%%%%%%%%%%%%%%%%%%%%%%%%%%%%%%%%%%%%%%%%%%%%%%%%%%%%%%%%%%%%%%%%%%%%%%%%%%%%%%%%
\begin{figure*}[ht!]
\centering
\includegraphics[width=0.85\linewidth]{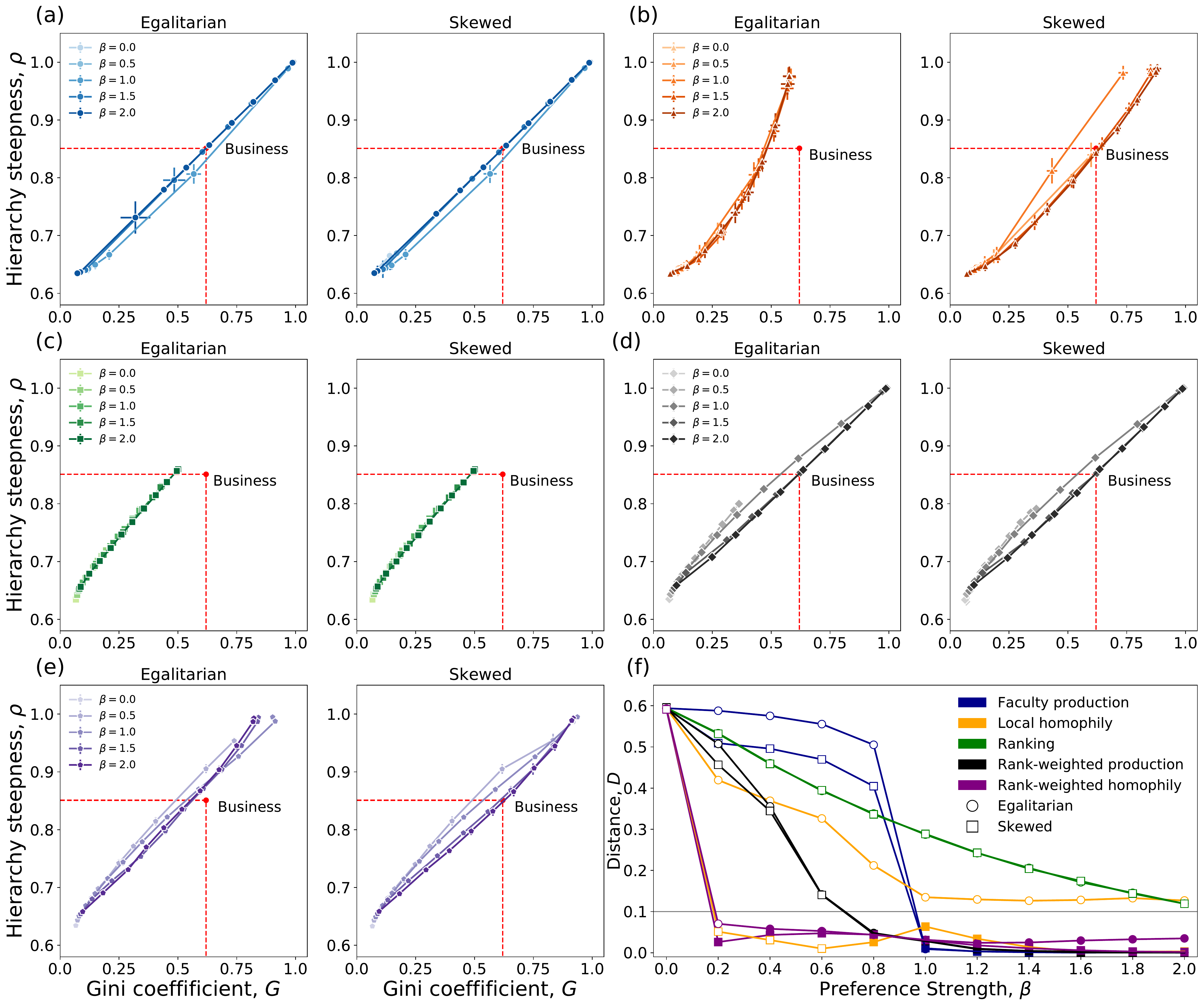}\\
\caption{{\bf Structural inequalities of five hiring models vs. empirical faculty hiring network of Business departments.} (a-e) Steady-state values of Gini coefficient $G$ and  hierarchy steepness $\rho$, for the five hiring models and two initial conditions for different choices of $\beta$ (for simplicity, we only show $\beta \in \{0.0, 0.5, 1.0, 1.5, 2.0\}$). Each line is a parametric plot in which we vary the randomness parameter over $0\leq p \leq 1$, and hence all models and choices of $\beta$ converge to the same point in inequality-space as $p\rightarrow 1$. Red dashed lines show the empirical values of $G$ and $\rho$ for the Business faculty hiring network. Results are averaged over $50$ simulations, and error bars indicate standard deviations. (f) Euclidean distance between an averaged structural inequality ($G$, $\rho$) obtained from simulation and their pair of empirical values for the Business hiring network. Filled markers represent $\beta$ values that can reproduce $G$ and $\rho$ with distance $D \leq 0.1$; open markers indicate $D > 0.1$ or a model is not able to reproduce the empirical inequality.}
\label{fig:ginirho}
\end{figure*}
%%%%%%%%%%%%%%%%%%%%%%%%%%%%%%%%%%%%%%%%%%%%%%%%%%%%%%%%%%%%%%%%%%%%%%%%%%%%%%%%%%%%%%%%%%

First, we found that fully or mostly random hiring ($p>0.5$) never reproduced the Gini coefficients $G$ and hierarchy steepness scores $\rho$ observed in empirical networks, irrespective of whether the initial conditions were egalitarian or skewed (Fig.~\ref{fig:ginirho}). This observation serves as a useful check by ruling out the possibility that the process of ranking nodes, in and of itself, leads to the false discovery of hierarchical structure in fully or mostly random networks.

Second, among the five proposed hiring preference functions, we found that the Local homophily model (egalitarian initialization) and Ranking model (either initialization) were unable to produce realistic networks, as measured by the distance from empirical $G$ and $\rho$ to those of the modeled networks, under any choice of preference strength $\beta$ or randomness $p$ (Fig.~\ref{fig:ginirho}f, Table~\ref{tab:error:BS}). The remaining models were able to reproduce realistic networks only for sufficiently large values of $\beta$ (Business, Fig.~\ref{fig:ginirho}; Computer Science, Fig.~\ref{Sfig:CS_ginirho}; History, Fig.~\ref{Sfig:HS_ginirho}).

To more precisely evaluate the plausibility of the remaining models, we identified combinations of preference strength $\beta$ and randomness $p$ that led each model and initial condition pair to produce plausibly realistic hiring networks (Tables~\ref{Stab:bestp}, \ref{sup_tab:CS_bestp}, and \ref{sup_tab:HS_bestp}; see Methods).~This exercise revealed that as preference strength increases, the properties of empirical hiring networks can be reproduced only if the probability of a random hire $p$ also increases. 
For instance, the Faculty production model with $\beta= 1.0$ reproduced empirical network inequalities at $p\approx0.007$, yet when $\beta$ increased to $2.0$ an increase in randomness to $p \approx 0.4$ was also required (Tables~\ref{Stab:bestp}, \ref{sup_tab:CS_bestp}, and \ref{sup_tab:HS_bestp}).

This coupling of $\beta$ and $p$ reflects a natural tension between the two parameters that the coupling balances.~That is, as $\beta$ increases, the hiring function concentrates more placement power among the most prestigious institutions, which tends to create hierarchies that are too steep and too unequal compared to the data. But, increasing $p$ balances this tendency by redistributing placement power equally among all institutions, which moderates the effect of $\beta$. 

%%%%%%%%%%%%%%%%%%%%%%%%%%%%%%%%%%%%%%%%%%%%%%%%%%%%%%%%%%%%%%%%%%%%%%%%%%%%%%%%%%%%%

%%%%%
%%%%%%%%%%%%%%%%%%%%%%%%%%%%%%%
\begin{table*}[!ht]
\setlength{\tabcolsep}{8pt}
\renewcommand{\arraystretch}{1.2}
\centering
\begin{tabular}{*{6}{c}|}
\textbf{Hiring Model} & \textbf{Init.} &\multicolumn{1}{c}{\textbf{\begin{tabular}[c]{@{}c@{}}Eigenvector centrality ($log$)\\ (Empirical, -6.71)\end{tabular}}} 
& 
\multicolumn{1}{c}{\textbf{\begin{tabular}[c]{@{}c@{}}Harmonic centrality\\ (Empirical, 0.53)\end{tabular}}} 
& 
\multicolumn{2}{c}{\textbf{\begin{tabular}[c]{@{}c@{}}Geodesic distance\\ (Empirical, 0.31)\end{tabular}}} \\ 
\hline
\multirow{2}{*}{\textbf{\begin{tabular}[c]{@{}c@{}}Faculty\\ production\end{tabular}}} 
& E  & $-2.69 \pm 4.21$  & $\mathbf{0.6 \pm 0.09}$  & \multicolumn{2}{c}{$\mathbf{0.4 \pm 0.11}$} \\ \cline{2-6} 
& S  & $-2.44 \pm 4.41$ & $0.62 \pm 0.1$& \multicolumn{2}{c}{$0.42 \pm 0.13$}  \\ 
\hline
\multirow{2}{*}{\textbf{\begin{tabular}[c]{@{}c@{}}Local \\ homophily\end{tabular}}}   & E  & --- & --- & \multicolumn{2}{c}{---}     \\ \cline{2-6}  & S              & $\mathbf{-7.49 \pm 3.9}$                                                                                         & $0.26 \pm 0.33$                                                                                           & \multicolumn{2}{c}{$\mathbf{0.4 \pm 0.13}$}                                                            \\ \hline
\multirow{2}{*}{\textbf{Ranking}}                                                      & E              & {---} &  {---} & \multicolumn{2}{c}{---}                                                                                                                  \\\cline{2-6} & S & --- & --- & \multicolumn{2}{c}{---} \\ \hline
\multirow{2}{*}{\textbf{\begin{tabular}[c]{@{}c@{}}Rank-weighted\\ production\end{tabular}}}   & E              & $-1.49 \pm 5.23$                                                                                                 & $0.67 \pm 0.16$                                                                                           & \multicolumn{2}{c}{$0.55 \pm 0.25$}                                                                    \\ \cline{2-6} 
                                                                                       & S              & $-1.51 \pm 5.21$                                                                                                 & $0.67 \pm 0.16$                                                                                           & \multicolumn{2}{c}{$0.55 \pm 0.25$}                                                                    \\ \hline
\multirow{2}{*}{\textbf{\begin{tabular}[c]{@{}c@{}}Rank-weighted\\ homophily\end{tabular}}}    & E              & $-1.48 \pm 5.25$                                                                                                 & $0.6 \pm 0.17 $                                                                                           & \multicolumn{2}{c}{$0.6 \pm 0.3$}                                                                      \\ \cline{2-6} 
                                                                                       & S              & $-1.48 \pm 5.25$                                                                                                 & $0.6 \pm 0.17$                                                                                            & \multicolumn{2}{c}{$0.59 \pm 0.29$}                                                                    \\ \hline
\end{tabular}
\caption{\textbf{Mean value of network summary statistics for five hiring models vs. the empirical values for Business.} Mean values of the three network summary statistics---eigenvector centrality, harmonic centrality, and normalized mean geodesic distance---for networks produced by each of the five hiring models vs. the corresponding values of the empirical Business network. Boldface values are close to the empirical value. Uncertainty indicates the average absolute distance between predicted and empirical value. For this comparison, parameter combinations achieving the closest structural inequality are applied when $\beta = 1.0$ (Table~\ref{Stab:bestp}). Two initializations (Init.) of Egalitarian (E) and Skewed (S) are used. Dashed line represents the Ranking and Local homophily model are unable to reproduce realistic structural inequalities.  All values are rounded to the second digit.}
\label{tab:error:BS}
\end{table*}
%%%%%%%%%%%%%%%%%%%%%%%%%%%%%%%%%%%%%%%%%%%%%%%%%%%%%%%%%%%%%%%%%%%%%%%%%%%%%%%%%%%%%%%%%%
%%%%%%%%%%%%%%%%%%%%%%%%%%%%%%%%%%%%%%%%%%%%%%%%%%%%%%%%%%%%%%%%%%%%%%%%%%%%%%%%%%%%%%%%%%

\begin{figure*}[ht!]
\includegraphics[width=0.85\linewidth]{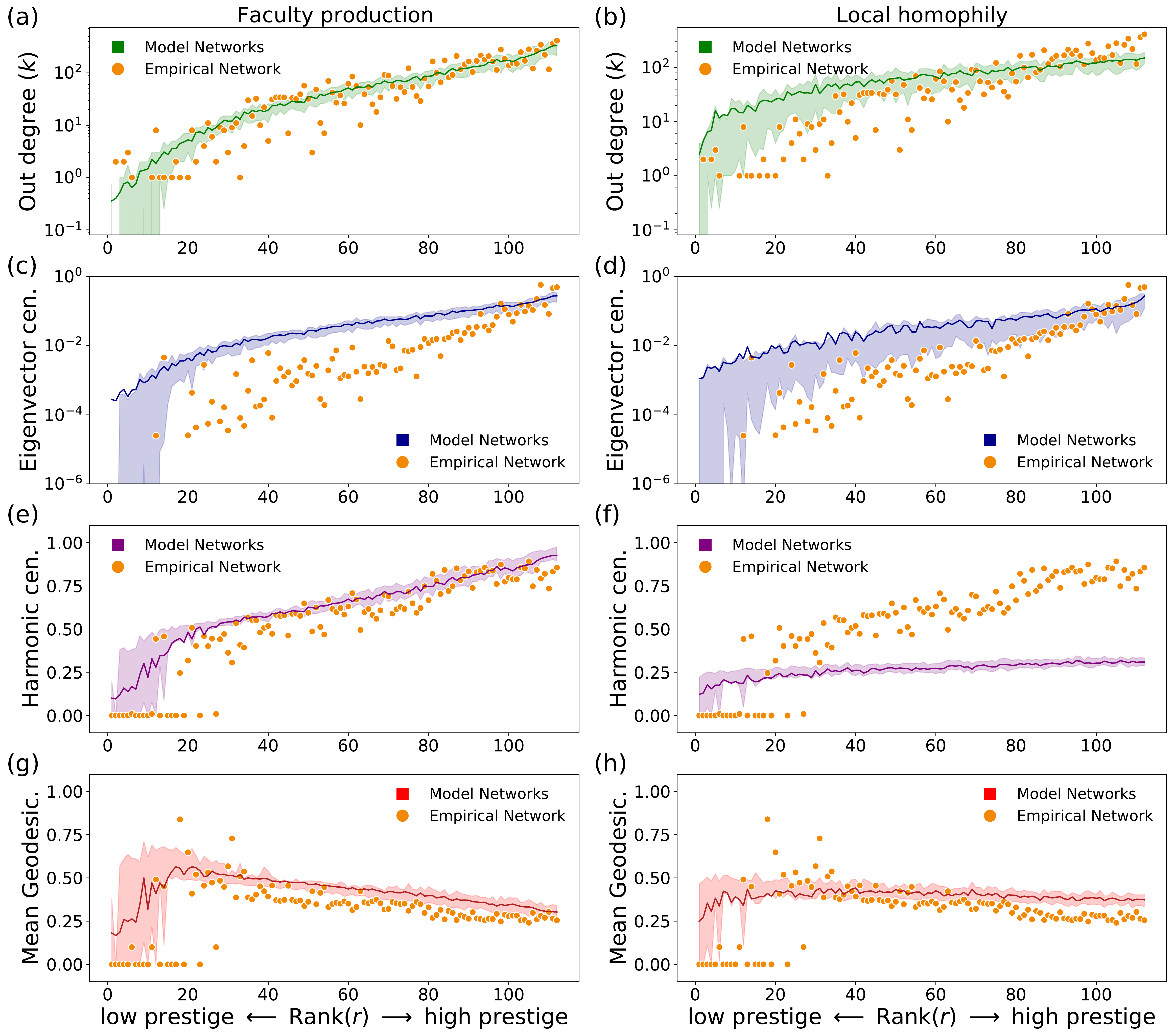}
\caption{\textbf{Structural patterns of hiring model networks for two mechanisms of Faculty production and Local homophily and the empirical Business hiring network as a function of prestige.} (a,b) Out-degree $k_i$, (c,d) eigenvector centrality, (e,f) harmonic centrality, and (g,h) mean geodesic distance normalized by network diameter. Panels on the left (a,c,e,g) show results for the Faculty production model, and panels on the right (b,d,f,h) show results for the Local homophily model (skewed initialization). Solid lines indicate the mean over $50$ simulations, and shaded region around the mean indicates the 25\% to 75\% quantile range. Orange points indicate the observed values for the Business hiring network. Large rank denotes higher prestige.}
\label{fig:networkcharacteristics:BS}
\end{figure*}

%%%%%%%%%%%%%%%%%%%%%%%%%%%%%%%%%%%%%%%%%%%%%%%%%%%%%%%%%%%%%%%%%%%%%%%%%%%%%%%%%%%%%%%%%%

\subsection*{Reproduction of Common Network Centralities} 
Seven of ten candidate hiring models can produce networks that replicate the observed measures of structural inequality in empirical hiring networks: This fact suggests they may all be plausible mechanisms, but raises the question of whether these models are also able to reproduce other statistical patterns in the empirical network's structure. To assess their ability to produce hiring networks that are realistic in other ways, we measured three additional network summary statistics as a function of node rank: eigenvector centrality, harmonic centrality, and normalized mean geodesic distance. In this analysis, we study networks generated by the best-fitting value of $p$ for $\beta=1$, identified in the previous section (Table~\ref{Stab:bestp}). 
This analysis provides a hard test for the models. While in the previous analysis, we fitted the parameters $\beta$ and $p$ in order to match the empirical values of the two structural inequality measures $G$ and $\rho$, here, models are scored by how well they reproduce network statistical patterns that were not part of estimating the model.

No single model was able to replicate all four centrality functions across all the fields. However, the Faculty production model with either initialization and the Local homophily model with skewed initialization produced the most realistic hiring networks.~While this fact can be seen by comparing empirical and model-produced summary statistics of centrality functions (Tables~\ref{tab:error:BS}, \ref{Stab:error:CS}, and \ref{Stab:error:HS}), scatter plots of centralities vs node rank reveal that the Faculty Production model further captures the more complex centrality-rank trends found in empirical data, followed by the Local homophily model with skewed initialization (Figs.~\ref{fig:networkcharacteristics:BS}, \ref{Sfig:networkcharacteristics:BS}, \ref{Sfig:networkcharacteristics:CS}, and \ref{Sfig:networkcharacteristics:HS}). The other models failed to meaningfully reproduce empirical centrality averages and centrality-rank trends. 

For both Faculty production and Local homophily models, the randomness parameter was low ($p=0.008$ and $0.007$; Table~\ref{Stab:bestp}), and yet the resulting simulated hiring networks nevertheless produce realistic measures of structural inequality and rank-centrality patterns.~These models’ agreement with the empirical rank-centrality patterns is notable, indicating that these mechanisms are able to reproduce additional structural patterns in faculty hiring networks that are not directly related to the fitted parameters of hierarchy steepness and faculty production inequality. This fact suggests that, generally speaking, even simple hiring preferences that reinforce the patterns of past hiring, either locally or globally, are capable of generating structurally plausible faculty hiring networks.

%%%%%%%%%%

%%%%%%%%%%%%%%%%%%%%%%%%%%%%%%%%

\begin{figure*}[ht!]
    \centering
    \includegraphics[width=0.85\linewidth]{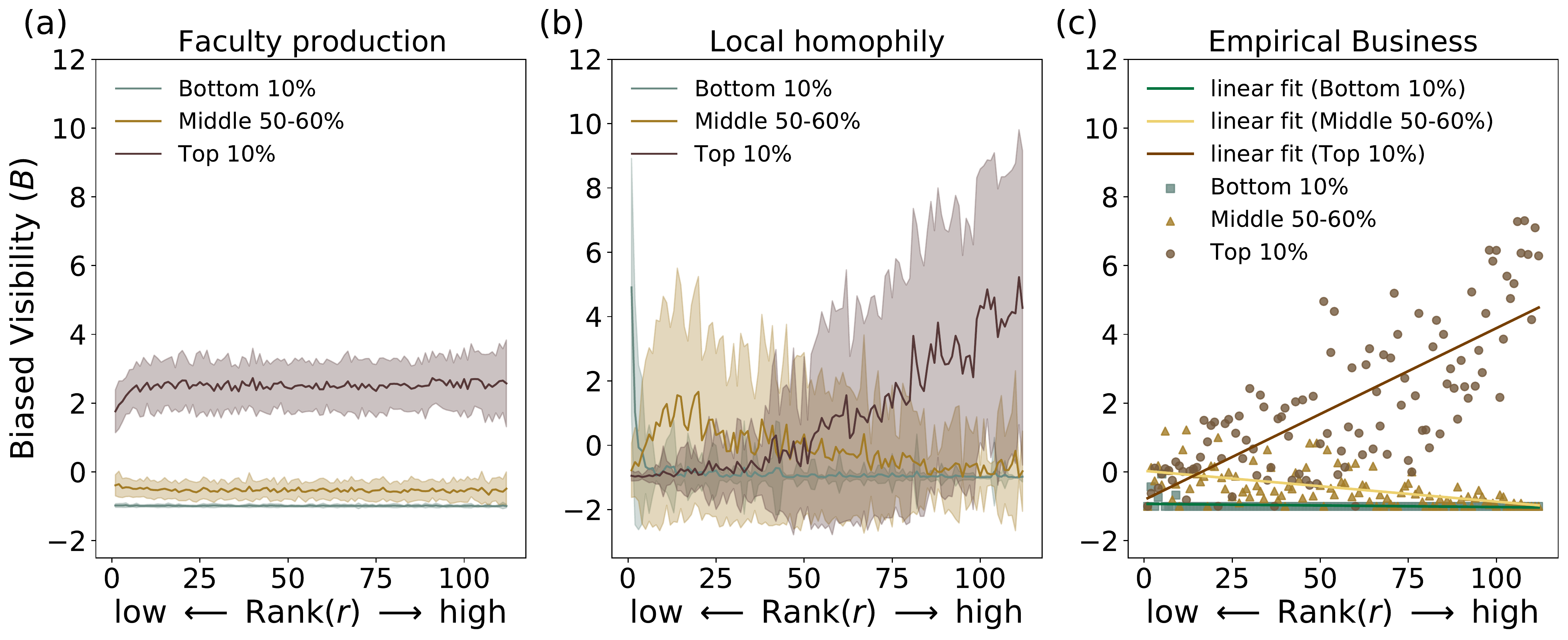}
    \caption{\textbf{Biased visibility $B$ for three groups of institutions (top 10\%, middle 50-60\%, and bottom 10\%) as a function of perceiver institution's rank $r$ for two models and the empirical Business network} (a) The Faculty production model, (b) Local homophily model, and (c) the empirical Business hiring network where each point is a department and lines indicate the best linear fit to the corresponding data. All models are from skewed initialization. Solid lines in (a,b) denote the average over $50$ networks ensembles, and the shaded region indicates the standard deviation. }
    \label{fig:visibility:BS}
\end{figure*}

\subsection*{Biased Visibility}
Hiring processes that explicitly reinforce the patterns of past hiring have the potential to also create stratified network structures in which a highly ranked node's neighbors are markedly different from those around a lower ranked node. As a result, some nodes may be far more visible to other nodes, and other nodes far less.~In the context of a faculty hiring network, biased visibility driven by prestige would imply that as the prestige of an institution grows, its visibility or adjacency to a wider variety of other institutions in the hiring network also grows.
To quantify this biased visibility, we measured the over- or under-representation of hires from the top-ranked 10\%, bottom-ranked 10\%, or 50-60th percentile in the in-neighborhoods of other nodes, arranged by rank.~We then compared the empirical and modeled patterns of biased visibility~\cite{mirta2012socialsampling, lerman2016Plosone} (see Eq.~\ref{eq:biasedvisibility} and Methods). 

Empirical faculty hiring networks show clear patterns of biased visibility by rank.~We find that low-prestige institutions are uniformly less visible to other institutions across the prestige hierarchy.~Institutions from the middle and top of the rankings are as visible as expected in the neighborhoods of low-ranked nodes, but as we consider a more prestigious subset of institutions, middle-ranked institutions become increasingly under-represented, while top-ranked institutions become overrepresented.~These patterns are consistently true across empirical Business (Figs.~\ref{fig:visibility:BS} and \ref{Sfig:bs_visibility_others}), Computer Science (Fig.~\ref{Sfig:CS_visibility}), and History networks (Fig.~\ref{Sfig:HS_visibility}), indicating that not only do the few most prestigious institutions produce the majority of all faculty, they also place these faculty broadly, across the entire prestige hierarchy.

In applying the same analysis to the simulated hiring networks produced by our best fitting models, we find that these prestige-correlated patterns of biased visibility were reproduced by the Local homophily model, but not by the Faculty production model (Figs.~\ref{fig:visibility:BS}, \ref{Sfig:bs_visibility_others}, \ref{Sfig:CS_visibility}, \ref{Sfig:HS_visibility}). This finding is further supported by Pearson correlations for top-ranked institutions between empirical and model-derived biased visibility for Business, Computer Science, and History, respectively: 0.73, 0.81, and 0.7 for the Local homophily model, vs 0.3, 0.45, and 0.5 for the Faculty production model (all Pearson correlation $p$-values $<0.001$). Thus, despite the fact that the Faculty production model better captures centrality-rank patterns, the Local homophily model better captures prestige-visibility patterns.~This discrepancy across measures of the networks' statistical structure suggests that the true mechanisms that explain real faculty hiring may reflect aspects of both models.

\subsection*{Rank Mobility}
In real academic systems, institutions may seek to make strategic choices in order to improve their ranking or prestige over time.~However, the mechanisms that drive these rankings, such as those studied here, may either mitigate or amplify the effects of individual choices, leading to a more or less dynamically stable hierarchy. A key advantage of our hiring models is the ability to study the dynamical consequences of hiring over time, and hence to investigate the long-term stability or fragility under natural dynamics. 

To quantify the natural drift in rankings over time, we first allow simulations to reach a steady state (see Methods).~We then simulated $M$ sequential retirements and hires, where $M$ is the total number of professors in the field (i.e., the total number of edges in the hiring network $M=\sum_{ij} A_{ij}$). Hence, we study a perturbation of the steady-state system in which roughly every faculty member is replaced once.~We record the initial steady-state rank quintile of each institution, and the corresponding final prestige quintile. This coordinate pair of initial and final quantile provides a simple measure of inter-generational rank mobility and drift.

\begin{figure}[ht!]
    \centering
    \includegraphics[width=0.99\linewidth]{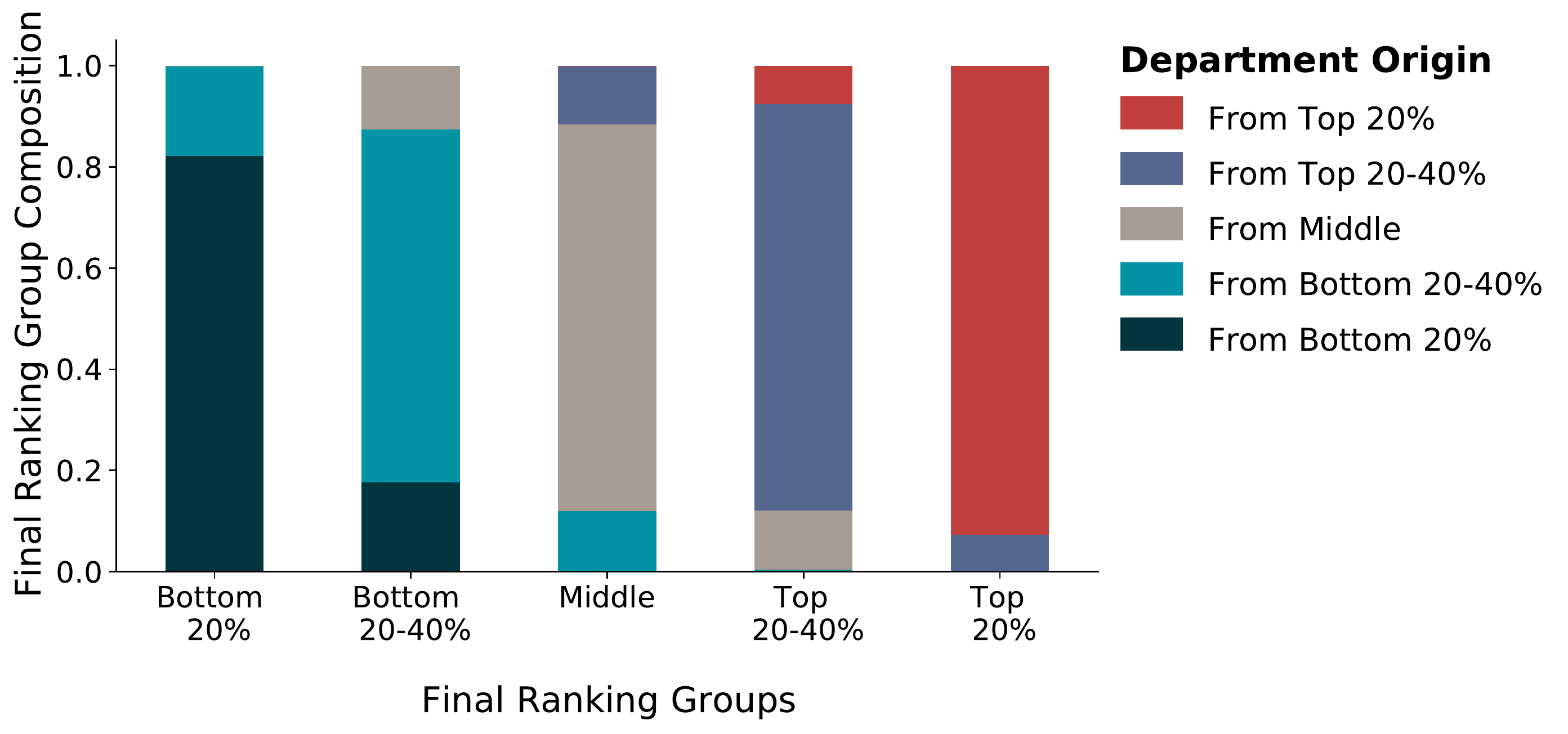}
    \caption{\textbf{Final ranking group composition of Business departments.} Rank mobility results derived from the best fitting Faculty production model with egalitarian initialization, with $\beta = 1.0$ and $p = 0.007$. The composition of departments' origin is traced and averaged over $50$ hiring simulated networks.}
    \label{fig:robust}
\end{figure}

Overall, we find that rank mobility under our best-fitting models is low. After a complete turnover of the Business faculty, on average, only 20\% of institutions had moved from one quintile to another, and the overwhelming majority moved only to an adjacent quintile (Fig.~\ref{fig:robust}). At the top of the rankings, 93\% of the institutions that were located in the final top quintile had also been there at the beginning of the experiment, with 82\% of institutions similarly staying in the bottom quintile (Fig.~\ref{fig:robust}). Similar results held for Computer Science (Fig.~\ref{Sfig:cs_robust}) and History (Fig.~\ref{Sfig:hs_robust}) simulations.

These findings suggest that once an institution is positioned near the top of the network hierarchy, prestige mobility by chance alone is limited. Only 7\%, 12\% and 8\% of institutions in the final top quintile of Business, Computer Science, and History networks, respectively, had drifted up from the quintile below (Faculty production model, egalitarian initialization). On the other hand, mobility into and out of the middle quintile was far greater, with new entrants to the middle quintile comprising 20-30\%, 29-49\%, and 23-52\% of middle-quintile institutions in Business, Computer Science, and History, respectively, after one generation. 

%%%%%%%%%%

\section*{Discussion}
Faculty hiring plays a fundamental role in shaping the composition of the scientific workforce~\cite{clauset2015sciadvan, morgan_SES}, and hence in determining the demographic composition of scientists~\cite{morgan_SES}, and what and whose particular scientific questions are studied~\cite{morgan2018EPJB}. Although past work has shown that doctoral prestige is predictive of whose graduates are hired as faculty~\cite{clauset2015sciadvan, samway2016www, DeBacco2018Sciadv}, we have lacked a deeper understanding the dynamics of and mechanisms that create and maintain prestige hierarchies~\cite{Larremore_2021_PNAS, Matthew_1968_Science, jobmarketsignaling_1973_spence, evolprestige_2000, Burris_academic_cste}. Without this understanding, the long-term stability of existing academic prestige hierarchies remains unclear—Would the same hierarchy re-emerge if we ``reset” the system? Would a different, but no less hierarchical organization emerge? How robust are the current hierarchies to perturbations? How much do positions within the hierarchy change under ``natural” dynamics? What structural models of ``prestige” can explain the emergence of the hierarchies we see today? Answers into these questions would help characterize the positive and negative roles of hierarchies in driving scientific discoveries, and inform efforts to make science a more equitable social endeavor.

Our investigation of five network-based models of faculty hiring dynamics sheds light on the general mechanisms that shape faculty hiring networks. Each of these models formalizes a different notion of institutional prestige, and hence allows us to investigate a variety of network questions about the emergence and maintenance of prestige hierarchies, and assess which notions of prestige produce the most realistic hierarchies. 

A key variable in our investigation was the out-degree distribution of the initial network, either egalitarian (all institutions starting at a similar position) or skewed (some institutions already ``ahead”), which allowed us to assess whether the emergence of a realistic hierarchy in faculty hiring depends on the system’s initial conditions. Across the five mechanisms, these initial conditions usually made little difference, and most mechanisms still produced realistic hierarchies for some choices of parameters. 

Using two standard measures of hierarchical structure, hierarchy steepness (fraction of hires ``down” the hierarchy) and inequality in faculty production (Gini coefficient in overall faculty production), to compare these five models, we find that two mechanisms were particularly successful at reproducing the characteristics of real faculty hiring networks. These successful mechanisms represent prestige dynamics that amplify the success of past hiring, either globally (Faculty production) or locally (Local homophily), as a kind of institution-level cumulative advantage.

Under two additional tests, we find that each of these two mechanisms outperforms each other in reproducing statistical patterns beyond those used to fit the models to the empirical data. The global mechanism (Faculty production) produces hiring networks with more realistic centrality scores as a function of position in the hierarchy, while the local mechanism (Local homophily) produces networks with more realistic measures of the unequal or biased visibility of nodes. These results suggest that the best explanation for the structure of real faculty hiring networks may be a mixture of global and local prestige, in which the likelihood of the graduate of institution $i$ being hired at $j$ depends both on how many faculty $i$ has placed anywhere and how many it has placed at $j$ in particular.

In our exploration of these models, each prestige-based mechanism was mixed with a non-prestige hiring mechanism, in which hires are chosen uniformly at random rather than by prestige.~Across our results, we find that strong hierarchies emerge even when only modest levels of prestige-based feedback in faculty hiring are used. However, we also found that the most realistic hierarchies required a trade-off between the strength of preference for prestige ($\beta$) and the tendency to hire independent of prestige ($p$). 

In both successful models for faculty hiring, we find that institution ranks are relatively stable over time, with the vast majority of institutions remaining in the same rank quintile after an entire generation of hiring (all faculty replaced once). In the absence of strategic hiring or placement behavior by individual institutions, the hiring system mixes poorly by rank. Hence, the prestige rankings embodied by hierarchical and unequal faculty hiring networks are unlikely to change within the careers of individual faculty without sustained and intentional strategy.

We note that the simplified mechanisms of faculty hiring we studied here explicitly omit any dependence on the characteristics of the individuals actually being hired. Studies examining the preferences of hiring committees suggest that additional factors, including individual productivity, gender, and postdoctoral experience, likely play important roles in their preferences for hiring different graduates~\cite{samway2016www,Willams_Nationalhiring_PNAS_2015, Headworth_cumulative_advan}. Nevertheless, the fact that our network-only models can accurately recapitulate the broad characteristics of empirical hiring networks suggests that, to a first approximation, the large-scale dynamics of hiring can be thought of as being mainly about prestige, which is known to correlate strongly with traditional metrics of ``merit."

All of the models introduced in this study are edge ``rewiring'' models, a broad class of generative network mechanisms in which the numbers of edges and vertices are held fixed, while edges are moved from one place in the network to another. This class of network formation processes is not well studied.~In contrast, much more is known about both network growth models, in which new nodes arrive steadily and form connections with existing nodes~\cite{Redner_uniformgrowth_PRL,Price_model_JASS}, and non-growth models, in which links are formed among a fixed set of nodes via conditionally independent draws~\cite{Erdos_model_1960,Holland_SBM_1983,Faust_SBM_1992, Newman_DSBM_PRE_2011}.~Our results hint at the richness of network dynamics that are possible from models that neither grow nor shrink, but instead rearrange edges dynamically. Such models are likely to find broad applications, and would benefit from the development of more general mathematical theory.

% Assumptions and limitations.
A number of our modeling assumptions may not hold true in practice, and these represent interesting directions of future work. For instance, real faculty hiring networks have neither a fixed number of nodes nor a fixed number of edges---within a single field, departments form, grow, shrink, and can even dissolve. Future work on the dynamics of faculty hiring should explore the causes, incentives, and thresholds underlying these dynamics, and their implications for the stability of hierarchies.~We also assumed that all institutions followed a universal preference function over candidates while in reality, these preferences may vary across institutions or across time. 
Finally, our models do not capture the two-sided nature of the hiring process, in which institutions make offers to individuals, and individuals choose which, if any, offers to accept. As a result, some job opening may go unfilled (no offers accepted) and some applicants may go unemployed (no offers received). A more realistic model of hiring would more explicitly capture these dynamics, which may lead to additional insights on the stability of hiring hierarchies, e.g., if the likelihood of an offer being received as accepted also correlates with prestige or the attributes of individuals. 

% Big Picture
Finally, our results bear on efforts to diversify the academy through changes in the mechanics of faculty hiring. Our models show that the self-reinforcing dynamics of prestige lead to entrenched hierarchies, which do not mix rapidly or dissipate on their own~\cite{Hanneman_2013_robust}. The robustness of prestige hierarchies has two key implications.~First, it suggests that those elite institutions that are currently responsible for placing the vast majority of faculty~\cite{clauset2015sciadvan} are likely to naturally remain in such positions unless preferences across the entire field change dramatically. Second, it suggests that these same institutions, which are in stable positions of high visibility and placement power, have the power to alter the demographic composition and research interests of the entire field, through their own PhD admissions and training.~Hence, although our findings predict that the dominant rank and placement power of these elite institutions is unlikely to change, their network positions may nevertheless enable them to rapidly reshape their fields in other significant ways.
\section*{Acknowledgements}
The authors thank Sam Zhang, Tzu-Chi Yen, Hunter Wapman, and Allison Morgan for helpful discussions. All authors acknowledge support from the Air Force Office of Scientific Research Award FA9550-19-1-0329.

% The \nocite command causes all entries in a bibliography to be printed out
% whether or not they are actually referenced in the text. This is appropriate
% for the sample file to show the different styles of references, but authors
% most likely will not want to use it.
\nocite{*}

\bibliography{apssamp}% Produces the bibliography via BibTeX.
\clearpage
\onecolumngrid
\appendix
\section*{Appendices}

%%%%%%%%%%%%%%%%%%%%%%%%%%%%%%%%%%%%%%%%
\renewcommand{\thefigure}{S\arabic{figure}}
\setcounter{figure}{0}
\renewcommand{\theequation}{S\arabic{equation}}
\setcounter{equation}{0}
\renewcommand{\thetable}{S\arabic{table}}
\setcounter{table}{0}

%%%%%%%%%%%%%%%%%%%%%%%%%%%%%%%%%%%%%%%%

\subsection*{Appendix A: Summary of the Empirical Hiring Networks}
\label{subsec:summary_networks}

%%%%%%-------------------%%%%%%%%%%%%
% \ignorespacesafterend
\begin{table*}[!ht]
\centering
\renewcommand{\arraystretch}{1.5}
\begin{tabular}{c|c|c|c}
                                    & \textbf{Business} & \textbf{Computer Science} & \textbf{History}      \\ \hline
\textbf{The number of institutions} & 112                    & 205                            & 144                        \\ \hline
\textbf{Mean institution's size}    & 83                     & 25                             & 32                         \\ \hline
\textbf{Full Professors}            & 4294 (46\%)            & 2400 (48\%)                   & 2097 (46\%)                \\ \hline
\textbf{Associate Professors}       & 2521 (27\%)            & 1772 (35\%)                    & 1611 (35\%)                \\ \hline
\textbf{Assistant Professors}       & 2521 (27\%)             & 860 (17\%)                    & 848 (19\%)                 \\ \hline
\textbf{Collection Period}          & Mar. 2012 - Dec. 2012  &  May 2011 - Mar. 2012    & Jan. 2013 - Aug. 2013 \\ \hline
\end{tabular}
\caption{\textbf{Summary of the Hiring Networks of Three Departments} Data summary for the collected faculty hiring networks for three disciplines:  Business, Computer Science, and History. The number of institutions and institution's size regarding the faculty hiring has been preserved in five hiring models.}
\label{sup_tab:network_summary}
\end{table*}

\clearpage

% \FloatBarrier  

 \subsection*{Appendix B: Preference Strength $\beta$ and the Probability of Random Hiring $p$ for Structural Inequalities}
 \subsubsection*{}

\begin{figure*}[ht!]
\centering
\includegraphics[width=0.9\linewidth]{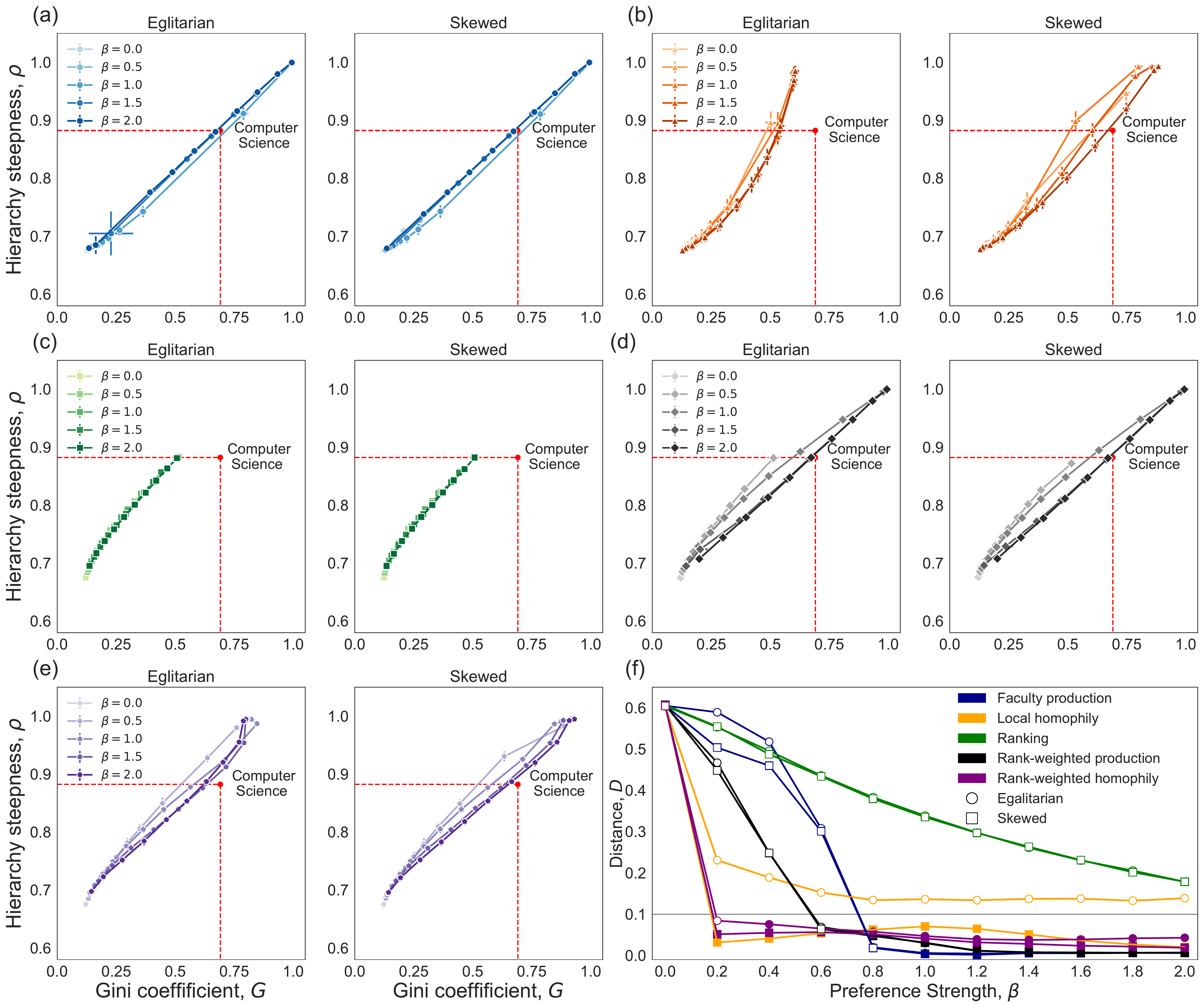}\\
\caption{\textbf{Best-fit values of the random hiring parameter $p$ for the Computer Science faculty hiring network}. (a-e) Steady-state values of Gini coefficient $G$ and  hierarchy steepness $\rho$, for the five hiring models and two initial conditions for different choices of $\beta$ (for simplicity, we only show $\beta \in \{0.0, 0.5, 1.0, 1.5, 2.0\}$). Each line is a parametric plot in which we vary the randomness parameter over $0\leq p \leq 1$, and hence all models and choices of $\beta$ converge to the same point in inequality-space as $p\rightarrow 1$. Red dashed lines show the empirical values of $G$ and $\rho$ for the Computer Science faculty hiring network. Results are averaged over $50$ simulations, and error bars indicate standard deviations. (f) Euclidean distance between an averaged structural inequality ($G$,$\rho$) obtained from simulation and their pair of empirical values for the Computer Science hiring network. Filled markers represent $\beta$ values that can reproduce $G$ and $\rho$ with distance $D \leq 0.1$; open markers indicate $D > 0.1$ or a model is not able to reproduce the empirical inequality.}
\label{Sfig:CS_ginirho}
\end{figure*}

\begin{figure*}[ht!]
\centering
\includegraphics[width=0.9\linewidth]{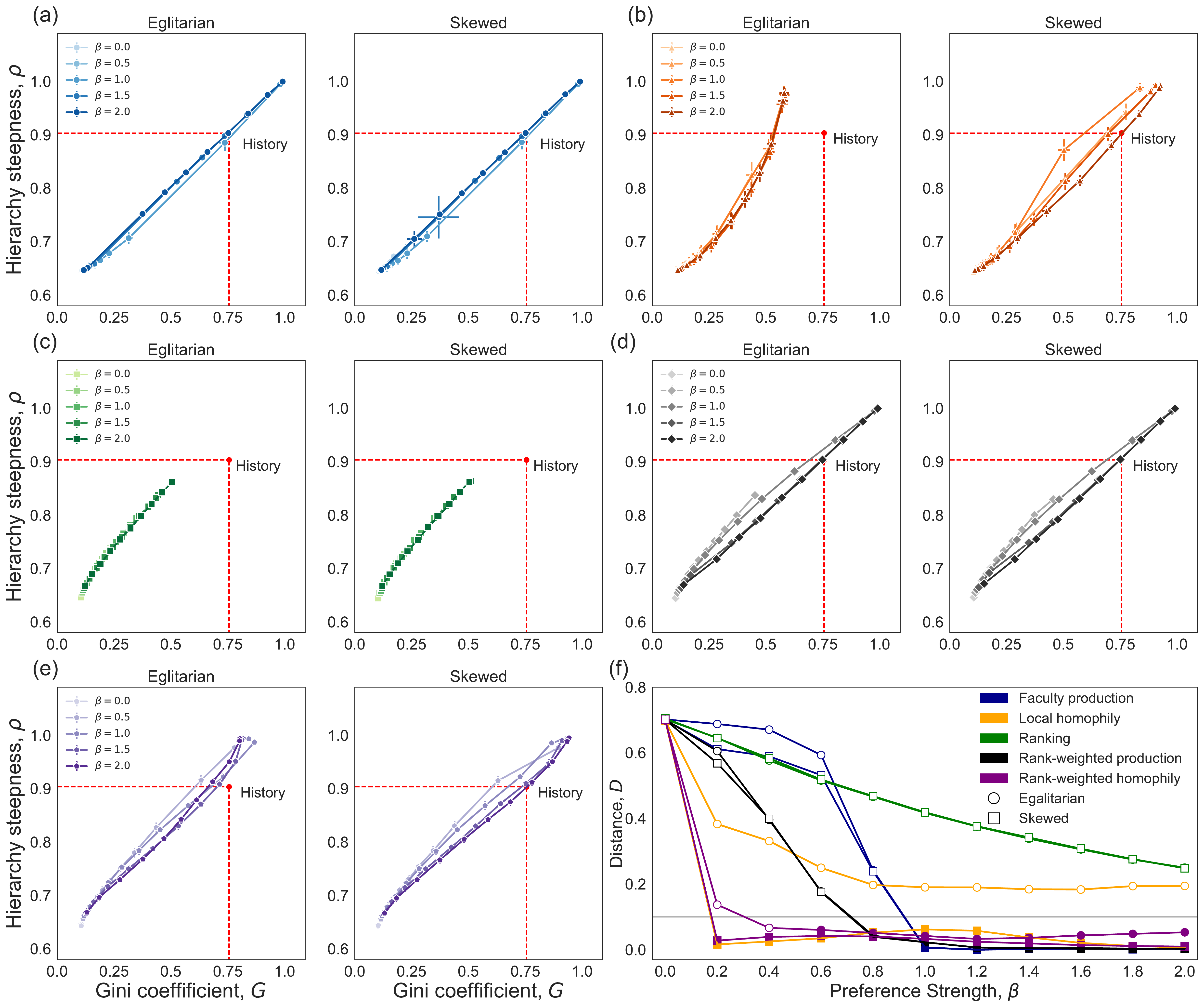}\\
\caption{\textbf{Best-fit values of the random hiring parameter $p$ for the History faculty hiring network}. (a-e) Steady-state values of Gini coefficient $G$ and  hierarchy steepness $\rho$, for the five hiring models and two initial conditions for different choices of $\beta$ (for simplicity, we only show $\beta \in \{0.0, 0.5, 1.0, 1.5, 2.0\}$). Each line is a parametric plot in which we vary the randomness parameter over $0\leq p \leq 1$, and hence all models and choices of $\beta$ converge to the same point in inequality-space as $p\rightarrow 1$. Red dashed lines show the empirical values of $G$ and $\rho$ for the History faculty hiring network. Results are averaged over $50$ simulations, and error bars indicate standard deviations. (f) Euclidean distance between an averaged structural inequality ($G$,$\rho$) obtained from simulation and their pair of empirical values for the History hiring network. Filled markers represent $\beta$ values that can reproduce $G$ and $\rho$ with distance $D \leq 0.1$; open markers indicate $D > 0.1$ or a model is not able to reproduce the empirical inequality.}
\label{Sfig:HS_ginirho}
\end{figure*}

\begin{table*}[!htbp]
\setlength{\tabcolsep}{8pt}
\renewcommand{\arraystretch}{1.2}
\centering
\begin{tabular}{c|c|c|c|c|c|c}
\multirow{2}{*}{\textbf{Hiring Model}}       & \multirow{2}{*}{\textbf{Init.}} & \multicolumn{5}{c}{\textbf{Preference Strength, \textbf{$\beta$}}}                                 \\ \cline{3-7} 
                                             &                                             & \textbf{0.0}   & \textbf{0.5}   & \textbf{1.0}   & \textbf{1.5}   & \textbf{2.0}   \\ \hline
\multirow{2}{*}{\textbf{Faculty production}} & E                                     & -              & -              & {0.007} & {0.39}  & {0.418} \\ \cline{2-7} 
                                             & S                                & -              & -              & {0.008} & {0.391} & {0.419} \\ \hline
\multirow{2}{*}{\textbf{Local homophily}}    & E                                     & -              & -              & -          & -          & -          \\ \cline{2-7} 
                                             & S                               & {-} & {0.001} & {0.007} & {0.28}  & {0.39}  \\ \hline
\multirow{2}{*}{\textbf{Ranking}}            & E                              & -              & -              & -              & -              & -          \\ \cline{2-7} 
                                             & S                              & -              & -              & -              & -              & 0.04          \\ \hline
\multirow{2}{*}{\textbf{Rank-weighted production}}   & E                                     & -              & -              & {0.247} & {0.418} & {0.429} \\ \cline{2-7} 
                                             & S                               & -              & -              & {0.244} & {0.418} & {0.423} \\ \hline
\multirow{2}{*}{\textbf{Rank-weighted homophily}}    & E                              & -              & {0.064}          & {0.236}          & {0.354}          & {0.342}          \\ \cline{2-7} 
                                             & S                              & {-} & {0.067} & {0.238} & {0.395} & {0.412} \\ \hline
\end{tabular}
\caption{\textbf{Best-fit values of the random hiring parameter $p$ for the Business faculty hiring network}. Each entry in the table presents the value of $p$ that minimized the distance between modeled networks and empirical networks for values of preference strength $\beta$, initializations (E, Egalitarian; S, Skewed) and hiring models, as shown. For combinations which were unable to plausibly reproduce the properties of empirical networks ($D>0.1$, see Methods), no value of $p$ is given.}
\label{Stab:bestp}
\end{table*}

\begin{table*}[ht!]
\centering
\setlength{\tabcolsep}{8pt}
\renewcommand{\arraystretch}{1.2}
\begin{tabular}{c|c|c|c|c|c|c}
\multirow{2}{*}{\textbf{Hiring Model}}       & \multirow{2}{*}{\textbf{Init.}} & \multicolumn{5}{c}{\textbf{Preference Strength, $\beta$}}                               \\ \cline{3-7} 
                                             &                                             & \textbf{0.0} & \textbf{0.5}   & \textbf{1.0}   & \textbf{1.5}   & \textbf{2.0}   \\ \hline
\multirow{2}{*}{\textbf{Faculty Pproduction}} & E                                      & -            & -              & {0.02}  & {0.383} & {0.407} \\ \cline{2-7} 
                                             & S                               & -            & -              & {0.02}  & {0.386} & {0.405} \\ \hline
\multirow{2}{*}{\textbf{Local homophily}}    & E                                     & -            & -              & -              & -              & -              \\ \cline{2-7} 
                                             & S                                & -            & {0.002} & {0.019} & {0.101} & {0.157} \\ \hline
\multirow{2}{*}{\textbf{Ranking}}            & E                                     & -            & -              & -              & -              & -              \\ \cline{2-7} 
                                             & S                                & -            & -              & -              & -              & -              \\ \hline
\multirow{2}{*}{\textbf{Rank-weighted production}}   & E                                      & -            & -              & {0.226} & {0.404} & {0.408} \\ \cline{2-7} 
                                             & S                               & -            & -              & {0.228} & {0.401} & {0.406} \\ \hline
\multirow{2}{*}{\textbf{Rank-weighted homophily}}    & E                                     & -            & {0.064} & {0.188} & {0.282} & {0.312} \\ \cline{2-7} 
                                             & S                               & -            & {0.066} & {0.187} & {0.285} & {0.311} \\ \hline
\end{tabular}
\caption{\textbf{Best-fit values of the random hiring parameter $p$ for the Computer Science faculty hiring network}. Each entry in the table presents the value of $p$ that minimized the distance between modeled networks and empirical networks for values of preference strength $\beta$, initializations (E, Egalitarian; S, Skewed) and hiring models, as shown. For combinations which were unable to plausibly reproduce the properties of empirical networks ($D>0.1$, see Methods), no value of $p$ is given.}
\label{sup_tab:CS_bestp}
\end{table*}

\begin{table*}[ht!]
\centering
\setlength{\tabcolsep}{8pt}
\renewcommand{\arraystretch}{1.2}
\begin{tabular}{c|c|c|c|c|c|c}
\multirow{2}{*}{\textbf{Hiring Model}}       & \multirow{2}{*}{\textbf{Init.}} & \multicolumn{5}{c}{\textbf{Preference Strength, $\beta$}}                               \\ \cline{3-7} 
                                             &                                             & \textbf{0.0} & \textbf{0.5}   & \textbf{1.0}   & \textbf{1.5}   & \textbf{2.0}   \\ \hline
\multirow{2}{*}{\textbf{Faculty production}} & E                                   & -            & -              & {0.009}  & {0.283} & {0.301} \\ \cline{2-7} 
                                             & S                              & -            & -              & {0.009}  & {0.281} & {0.3} \\ \hline
\multirow{2}{*}{\textbf{Local homophily}}    & E                                    & -            & -              & -              & -              & -              \\ \cline{2-7} 
                                             & S                                & -            & {0.001} & {0.007} & {0.092} & {0.155} \\ \hline
\multirow{2}{*}{\textbf{Ranking}}            & E                                 & -            & -              & -              & -              & -              \\ \cline{2-7} 
                                             & S                                & -            & -              & -              & -              & -              \\ \hline
\multirow{2}{*}{\textbf{Rank-weighted production}}   & E                                     & -            & -              & {0.161} & {0.3} & {0.301} \\ \cline{2-7} 
                                             & S                           & -            & -              & {0.161} & {0.298} & {0.302} \\ \hline
\multirow{2}{*}{\textbf{Rank-weighted homophily}}    & E                                      & -            & {0.031} & {0.134} & {0.215} & {0.226} \\ \cline{2-7} 
                                             & S                                & -            & {0.034} & {0.134} & {0.215} & {0.222} \\ \hline
\end{tabular}
\caption{\textbf{Best-fit values of the random hiring parameter $p$ for the History faculty hiring network}. Each entry in the table presents the value of $p$ that minimized the distance between modeled networks and empirical networks for values of preference strength $\beta$, initializations (E, Egalitarian; S, Skewed) and hiring models, as shown. For combinations which were unable to plausibly reproduce the properties of empirical networks ($D>0.1$, see Methods), no value of $p$ is given.}
\label{sup_tab:HS_bestp}
\end{table*}
\clearpage
%%%%%%%%%%%%%%%%%%%%%%%%%%%%%%%%%%%%%%%%%

\subsection*{Appendix C: Network Properties of Empirical Hiring Network}

\begin{figure*}[!ht]
\centering

\includegraphics[width=0.95\linewidth]{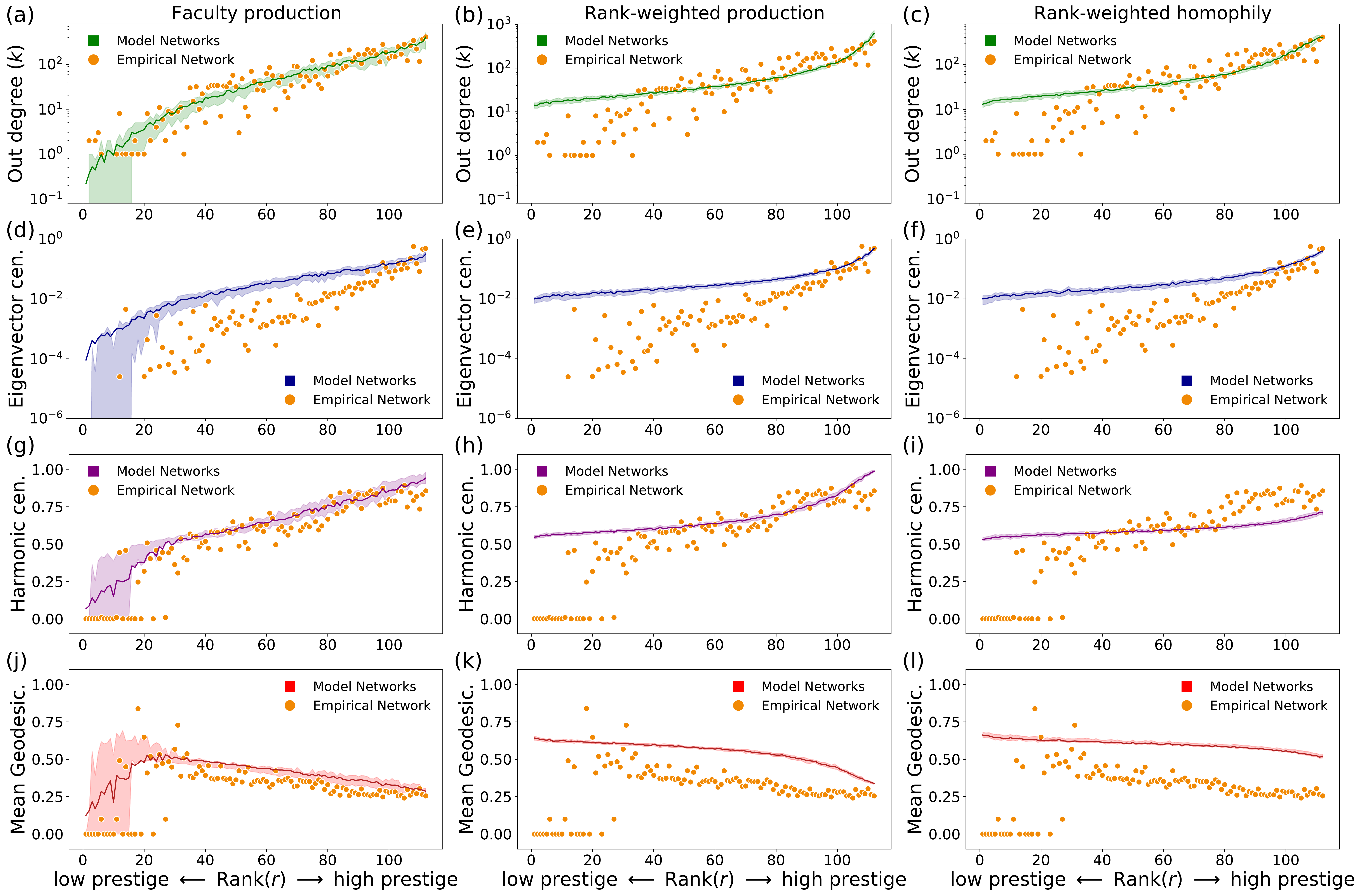}

\caption{\textbf{Structural patterns of hiring model networks for three mechanisms of Faculty production, Rank-weighted production, Rank-weighted homophily and the empirical Business hiring network as a function of prestige.} (a, b, c) Out-degree $k_i$, (d, e, f) eigenvector centrality, (g, h, i) harmonic centrality, and (j, k, l) mean geodesic distance normalized by network diameter. Panels on the left (a, d, g, j) show results for the Faculty production model, and panels on the middle (b, e, h, k) show results for the Rank-weighted production model. Panels on the right (c, f, i, l) show results for the Rank-weighted homophily model. All models simulated with egalitarian initialization. Solid lines indicate the mean over $50$ simulations, and shaded region around the mean indicates the 25\% to 75\% quantile range. Orange points indicate the observed values for the Business hiring network. Large rank denotes higher prestige.}
\label{Sfig:networkcharacteristics:BS}
\end{figure*}

\begin{table*}[!ht]
\setlength{\tabcolsep}{8pt}
\renewcommand{\arraystretch}{1.1}
\begin{tabular}{c|c|c|c|c|c}
\textbf{Hiring Model}                                                                  & \textbf{Init.} & \multicolumn{1}{c|}{\textbf{\begin{tabular}[c]{@{}c@{}}Eigenvector centrality ($log$)\\ (Empirical, -6.7)\end{tabular}}} & \multicolumn{1}{c|}{\textbf{\begin{tabular}[c]{@{}c@{}}Harmonic centrality\\ (Empirical, 0.34)\end{tabular}}} & \multicolumn{2}{c}{\textbf{\begin{tabular}[c]{@{}c@{}}Geodesic distance\\ (Empirical, 0.34)\end{tabular}}} \\ \hline
\multirow{2}{*}{\textbf{\begin{tabular}[c]{@{}c@{}}Faculty\\ production\end{tabular}}} & E              & $-3.85 \pm 3.5 $                                                                                                & $0.39 \pm 0.09$                                                                                           & \multicolumn{2}{c}{$\mathbf{0.34 \pm 0.11}$}                                                                    \\ \cline{2-6} 
                                                                                       & S              & $\mathbf{-3.89 \pm 3.52}$                                                                                                & $0.39 \pm 0.09$                                                                                           & \multicolumn{2}{c}{$0.35 \pm 0.12$}                                                                    \\ \hline
\multirow{2}{*}{\textbf{\begin{tabular}[c]{@{}c@{}}Local \\ homophily\end{tabular}}}   & E             & --- & ---&\multicolumn{2}{c}{---} \\\cline{2-6} 
                                                                                       & S              & $-13.88 \pm 8.4$                                                                                                & $0.17 \pm 0.21$                                                                                           & \multicolumn{2}{c}{$0.34 \pm 0.14$}                                                                    \\ \hline
\multirow{2}{*}{\textbf{Ranking}}                                       
& E  & --- & ---&\multicolumn{2}{c}{---} \\ \cline{2-6}
& S & --- & ---&\multicolumn{2}{c}{---} \\ \hline
\multirow{2}{*}{\textbf{\begin{tabular}[c]{@{}c@{}}Rank-weighted\\ production\end{tabular}}}   & E             & $-1.8 \pm 4.91$                                                                                                 & $0.45 \pm 0.12$                                                                                           & \multicolumn{2}{c}{$0.47 \pm 0.17$}                                                                    \\ \cline{2-6} 
                                                                                       & S              & $-1.79 \pm 4.91$                                                                                                & $0.45 \pm 0.12$                                                                                           & \multicolumn{2}{c}{$0.46 \pm 0.16$}                                                                    \\ \hline
\multirow{2}{*}{\textbf{\begin{tabular}[c]{@{}c@{}}Rank-weighted\\ homophily\end{tabular}}}    & E             & $-1.92 \pm 4.79$                                                                                                & $\mathbf{0.35 \pm 0.13}$                                                                                           & \multicolumn{2}{c}{$0.43 \pm 0.15$}                                                                    \\ \cline{2-6} 
                                                                                       & S           & $-1.96 \pm 4.76$                                                                                                & $\mathbf{0.35 \pm 0.13}$                                                                                           & \multicolumn{2}{c}{$0.44 \pm 0.16$}                                                                    \\ \hline
\end{tabular}
\caption{\textbf{Mean value of network summary statistics for five hiring models vs. the empirical values for Computer Science.} Mean values of the three network summary statistics---eigenvector centrality, harmonic centrality, and normalized mean geodesic distance---for networks produced by each of the five hiring models vs. the corresponding values of the empirical Computer Science network. Boldface values are close to the empirical value. Uncertainty indicates the average absolute distance between predicted and empirical value. For this comparison, parameter combinations achieving the closest structural inequality are applied when $\beta = 1.0$ (Table~\ref{sup_tab:CS_bestp}). Two initializations (Init.) of Egalitarian (E) and Skewed (S) are used. Dashed line represents the Ranking and Local homophily model are unable to reproduce realistic structural inequalities. All values are rounded to the second digit.}
\label{Stab:error:CS}
\end{table*}

\begin{table*}[!ht]
\centering
\setlength{\tabcolsep}{8pt}
\renewcommand{\arraystretch}{1.2}
% Please add the following required packages to your document preamble:
% \usepackage{multirow}
\begin{tabular}{c|c|c|c|c|c}
\textbf{Hiring Model}                                                                  & \textbf{Init.} & \multicolumn{1}{c|}{\textbf{\begin{tabular}[c]{@{}c@{}}Eigenvector centrality ($log$)\\ (Empirical, -5.32)\end{tabular}}} & \multicolumn{1}{c|}{\textbf{\begin{tabular}[c]{@{}c@{}}Harmonic centrality\\ (Empirical, 0.38)\end{tabular}}} & \multicolumn{2}{c}{\textbf{\begin{tabular}[c]{@{}c@{}}Geodesic distance\\ (Empirical, 0.27)\end{tabular}}} \\ \hline
\multirow{2}{*}{\textbf{\begin{tabular}[c]{@{}c@{}}Faculty\\ production\end{tabular}}} & E              & $\mathbf{-5.35 \pm 2.38}$                                                                                        & $\mathbf{0.38 \pm 0.09}$                                                                                  & \multicolumn{2}{c}{$0.29 \pm 0.15$}                                                                    \\ \cline{2-6} 
                                                                                       & S              & $-5.38 \pm 2.39$                                                                                                 & $0.37 \pm 0.09$                                                                                           & \multicolumn{2}{c}{$0.29 \pm 0.15$}                                                                    \\ \hline
\multirow{2}{*}{\textbf{\begin{tabular}[c]{@{}c@{}}Local \\ homophily\end{tabular}}}   & E  & --- & ---&\multicolumn{2}{c}{---} \\ \cline{2-6}                                                                                                                                                                                                                                                                 
                                                                                       & S              & $-44.37 \pm 39.06$                                                                                               & $0.12 \pm 0.28$                                                                                           & \multicolumn{2}{c}{$\mathbf{0.27 \pm 0.13}$}                                                           \\ \hline
\multirow{2}{*}{\textbf{Ranking}}                                       
& E  & --- & ---&\multicolumn{2}{c}{---} \\ \cline{2-6}
& S & --- & ---&\multicolumn{2}{c}{---} \\ \hline                         
\multirow{2}{*}{\textbf{\begin{tabular}[c]{@{}c@{}}Rank-weighted\\ production\end{tabular}}}   & E              & $-2.06 \pm 3.31$                                                                                                 & $0.49 \pm 0.13$                                                                                           & \multicolumn{2}{c}{$0.46 \pm 0.21$}                                                                    \\ \cline{2-6} 
                                                                                       & S              & $-2.08 \pm 3.3$                                                                                                  & $0.49 \pm 0.13$                                                                                           & \multicolumn{2}{c}{$0.46 \pm 0.2$}                                                                     \\ \hline
\multirow{2}{*}{\textbf{\begin{tabular}[c]{@{}c@{}}Rank-weighted\\ homophily\end{tabular}}}    & E              & $-2.28 \pm 3.09$                                                                                                 & $0.38 \pm 0.15$                                                                                           & \multicolumn{2}{c}{$0.45 \pm 0.2$}                                                                     \\ \cline{2-6} 
                                                                                       & S              & $-2.26 \pm 3.1$                                                                                                  & $0.38 \pm 0.15$                                                                                           & \multicolumn{2}{c}{$0.44 \pm 0.19$}                                                                    \\ \hline
\end{tabular}
\caption{\textbf{Mean value of network summary statistics for five hiring models vs. the empirical values for History.} Mean values of the three network summary statistics---eigenvector centrality, harmonic centrality, and normalized mean geodesic distance---for networks produced by each of the five hiring models vs. the corresponding values of the empirical History network. Boldface values are close to the empirical value. Uncertainty indicates the average absolute distance between predicted and empirical value. For this comparison, parameter combinations achieving the closest structural inequality are applied when $\beta = 1.0$ (Table~\ref{sup_tab:HS_bestp}). Two initializations (Init.) of Egalitarian (E) and Skewed (S) are used. Dashed line represents the Ranking and Local homophily model are unable to reproduce realistic structural inequalities. All values are rounded to the second digit.}
\label{Stab:error:HS}
\end{table*}

\begin{figure*}[!ht]
\centering

\includegraphics[width=0.95\linewidth]{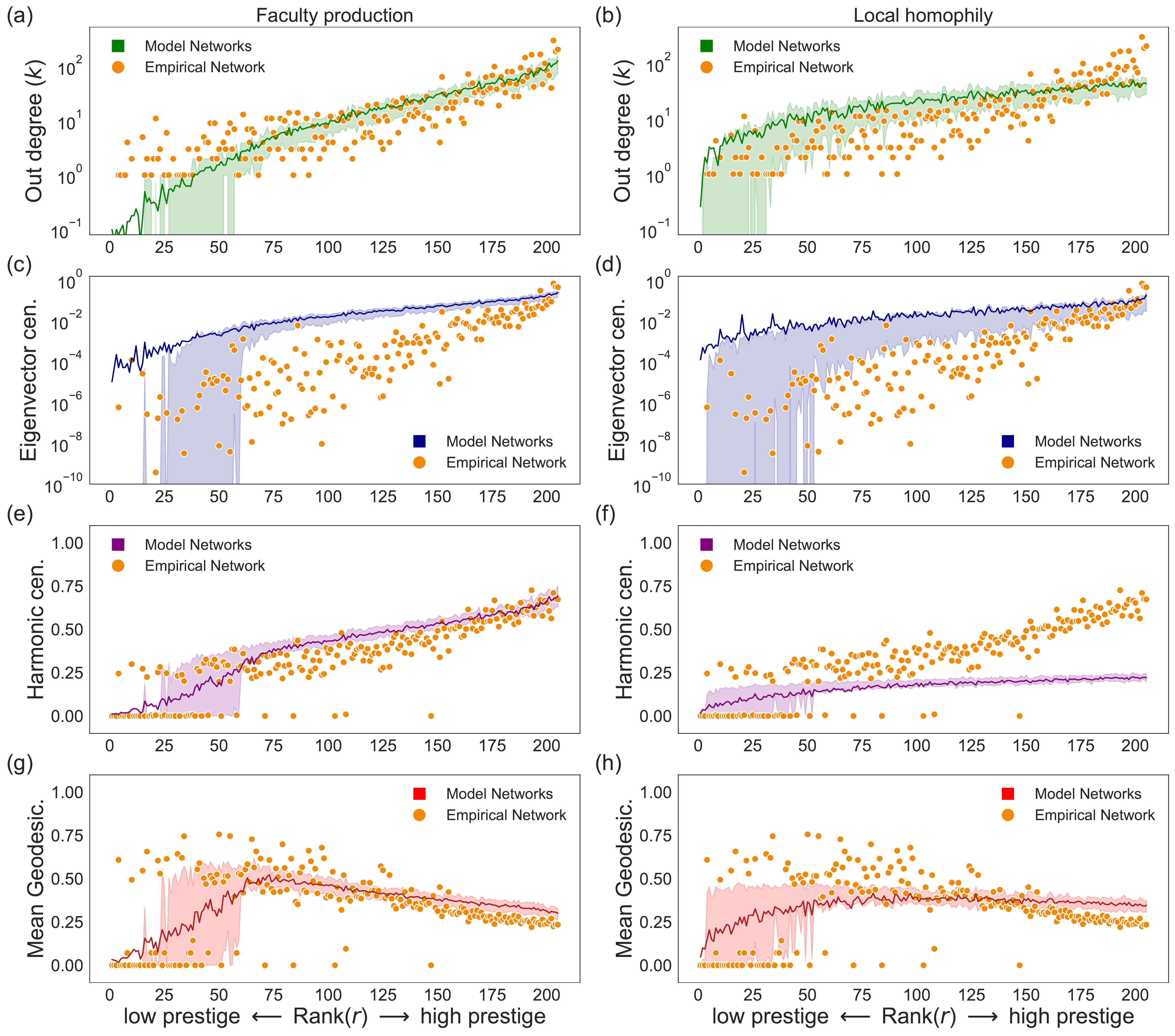}
\caption{\textbf{Structural patterns of hiring model networks for two mechanisms of Faculty production and Local homophily and the empirical Computer Science hiring network as a function of prestige.} (a, b) Out-degree $k_i$, (c, d) eigenvector centrality, (e, f) harmonic centrality, and (g, h) mean geodesic distance normalized by network diameter. Panels on the left (a, c, e, g) show results for the Faculty production model, and panels on the right (b, d, f, h) show results for the Local homophily model (skewed initialization). Solid lines indicate the mean over $50$ simulations, and shaded region around the mean indicates the 25\% to 75\% quantile range. Orange points indicate the observed values for the Computer Science hiring network. Large rank denotes higher prestige.}
\label{Sfig:networkcharacteristics:CS}
\end{figure*}

\begin{figure*}[!ht]
\centering
\includegraphics[width=0.95\linewidth]{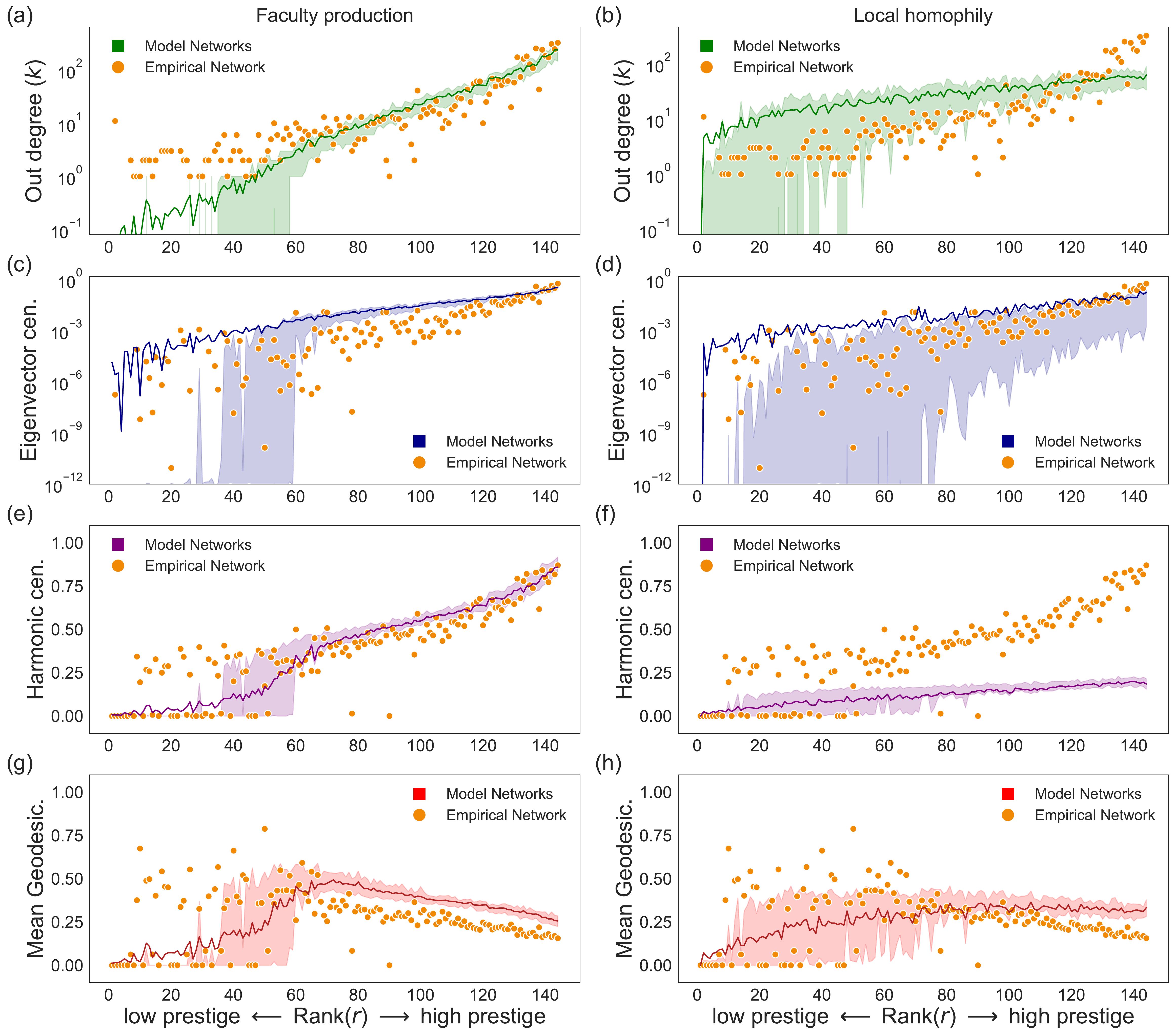}
\caption{\textbf{Structural patterns of hiring model networks for two mechanisms of Faculty production and Local homophily and the empirical History hiring network as a function of prestige.} (a, b) Out-degree $k_i$, (c, d) eigenvector centrality, (e, f) harmonic centrality, and (g, h) mean geodesic distance normalized by network diameter. Panels on the left (a, c, e, g) show results for the Faculty production model, and panels on the right (b, d, f, h) show results for the Local homophily model (skewed initialization). Solid lines indicate the mean over $50$ simulations, and shaded region around the mean indicates the 25\% to 75\% quantile range. Orange points indicate the observed values for the History hiring network. Large rank denotes higher prestige.}
\label{Sfig:networkcharacteristics:HS}
\end{figure*}

\clearpage

\subsection*{Appendix D: Biased Visibility from Hiring Network}

\begin{figure*}[!ht]
\centering

\includegraphics[width=0.95\linewidth]{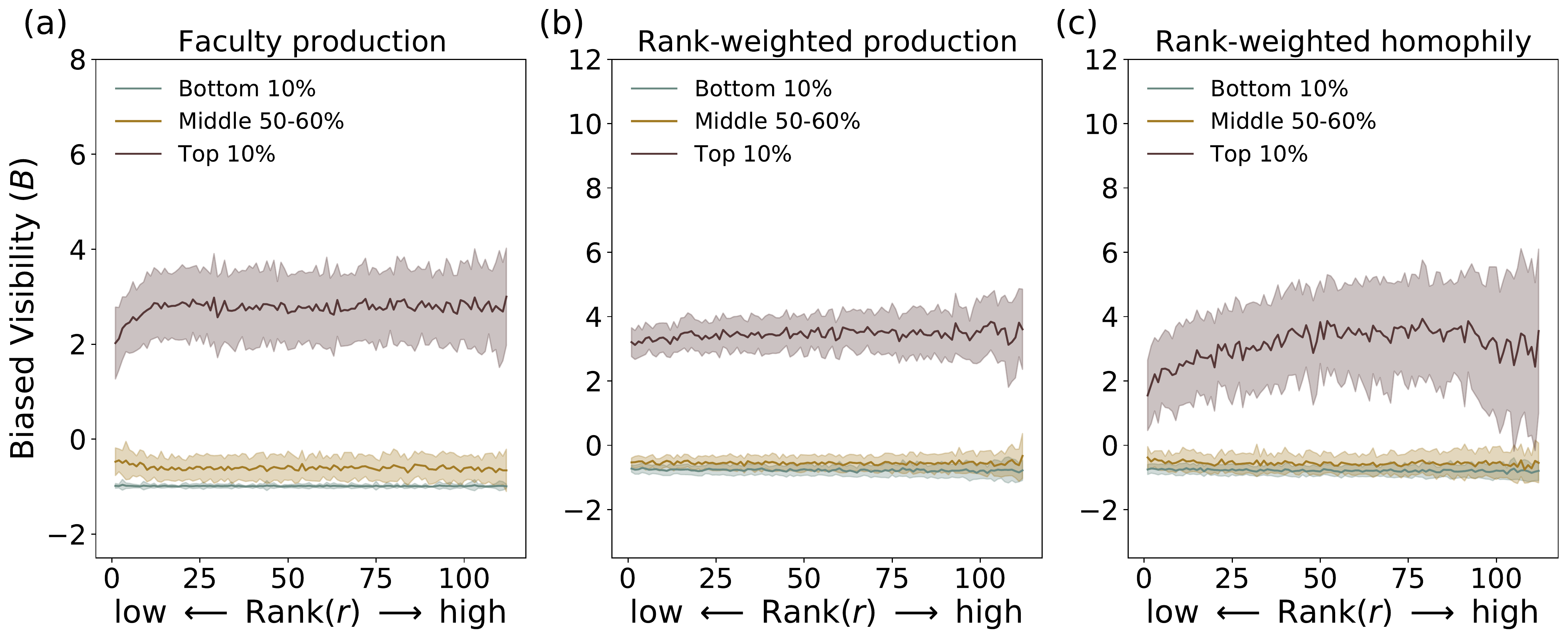}

\caption{\textbf{Biased visibility $B$ for three groups of institutions (top 10\%, middle 50-60\%, and bottom 10\%) as a function of perceiver institution's rank $r$ for three models for Business network} (a) Faculty production model, (b) Rank-weighted production model, and (c) Rank-weighted homophily. All models started from the egalitarian initialization. Solid lines in (a, b, c) denote the average over $50$ networks ensembles, and the shaded region indicates the standard deviation.}
\label{Sfig:bs_visibility_others}
\end{figure*}

% \FloatBarrier  

\begin{figure*}[ht!]
\centering

\includegraphics[width=0.95\linewidth]{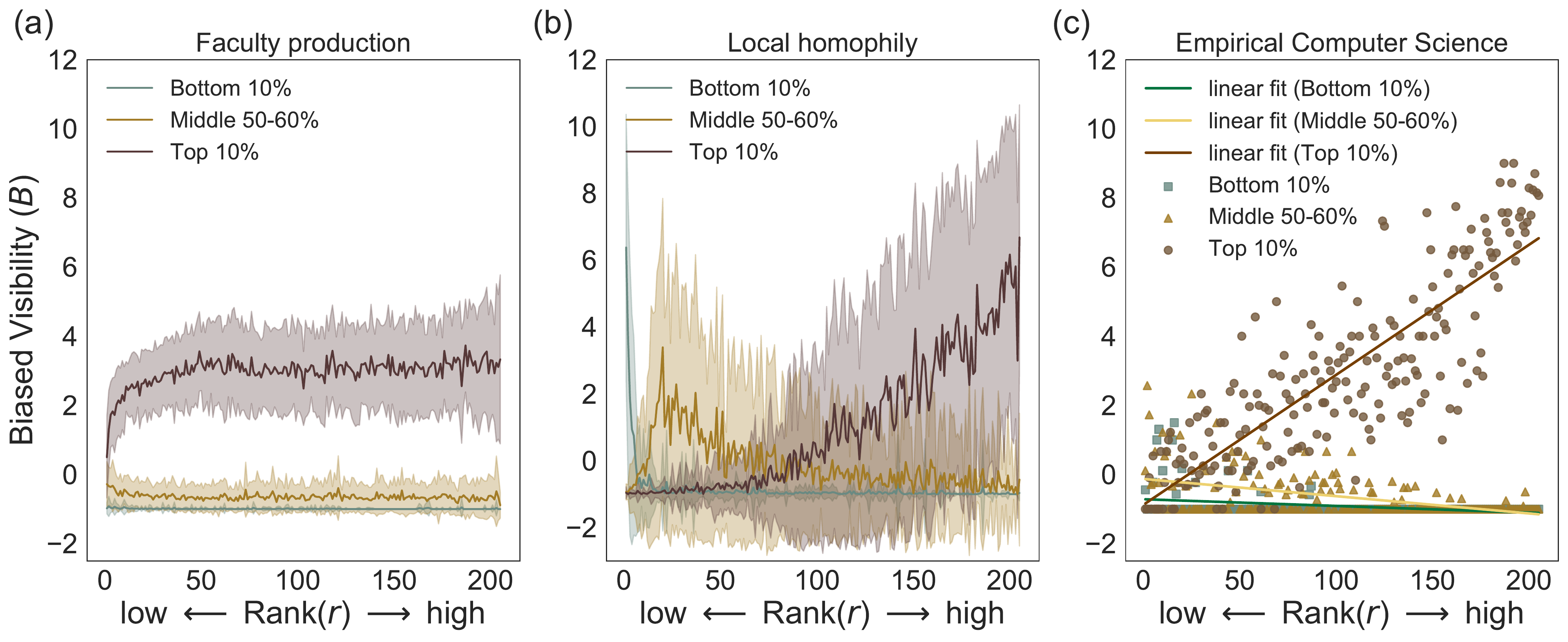}
\caption{\textbf{Biased visibility $B$ for three groups of institutions (top 10\%, middle 50-60\%, and bottom 10\%) as a function of perceiver institution's rank $r$ for two models and the empirical Computer Science network} (a) The Faculty production model, (b) Local homophily model, and (c) the empirical Computer Science hiring network where each point is a department and lines indicate the best linear fit to the corresponding data. All models are from skewed initialization. Solid lines in (a, b) denote the average over $50$ networks ensembles, and the shaded region indicates the standard deviation. }
\label{Sfig:CS_visibility}
\end{figure*}

\begin{figure*}[ht!]
\centering
\includegraphics[width=0.95\linewidth]{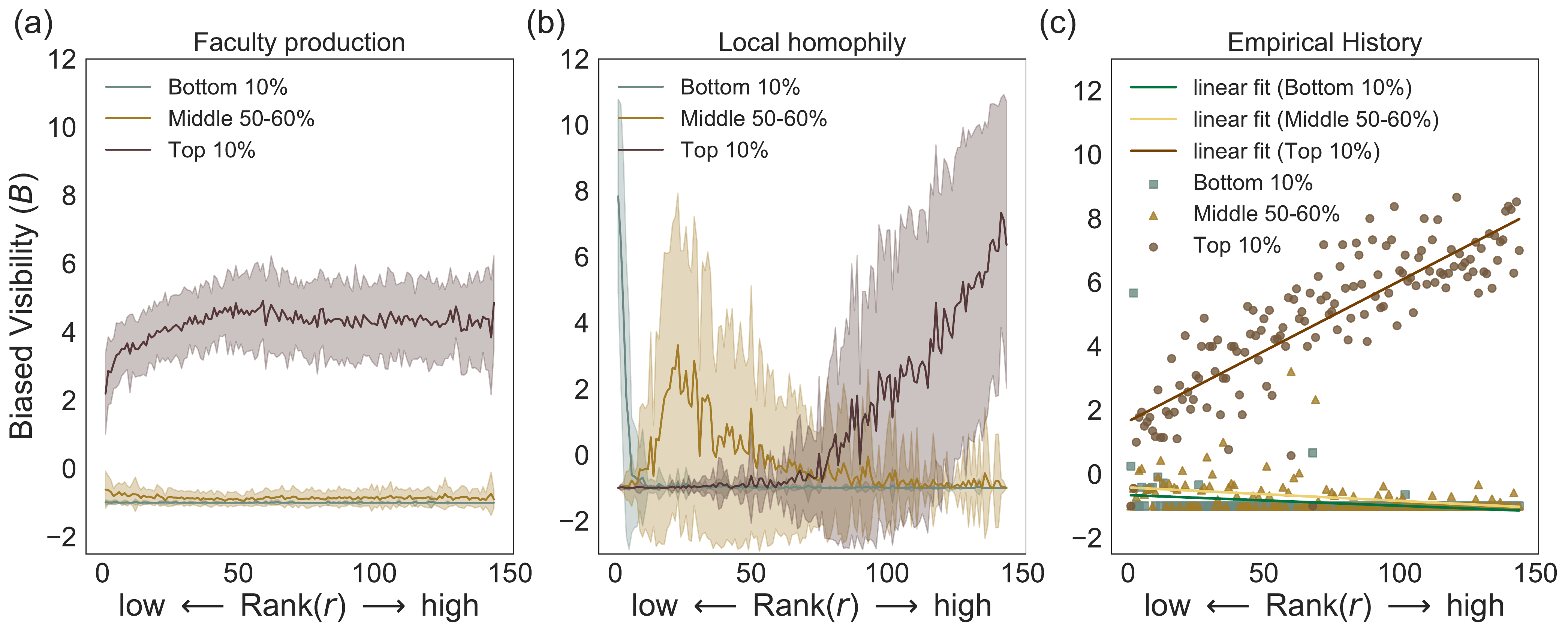}
\caption{\textbf{Biased visibility $B$ for three groups of institutions (top 10\%, middle 50-60\%, and bottom 10\%) as a function of perceiver institution's rank $r$ for two models and the empirical History network} (a) The Faculty production model, (b) Local homophily model, and (c) the empirical History hiring network where each point is a department and lines indicate the best linear fit to the corresponding data. All models are from skewed initialization. Solid lines in (a, b) denote the average over $50$ networks ensembles, and the shaded region indicates the standard deviation.}
\label{Sfig:HS_visibility}
\end{figure*}

\clearpage

\subsection*{Appendix E: Stability of Rank of Hiring Network}

\begin{figure*}[ht!]
\centering
\includegraphics[width=0.85\linewidth]{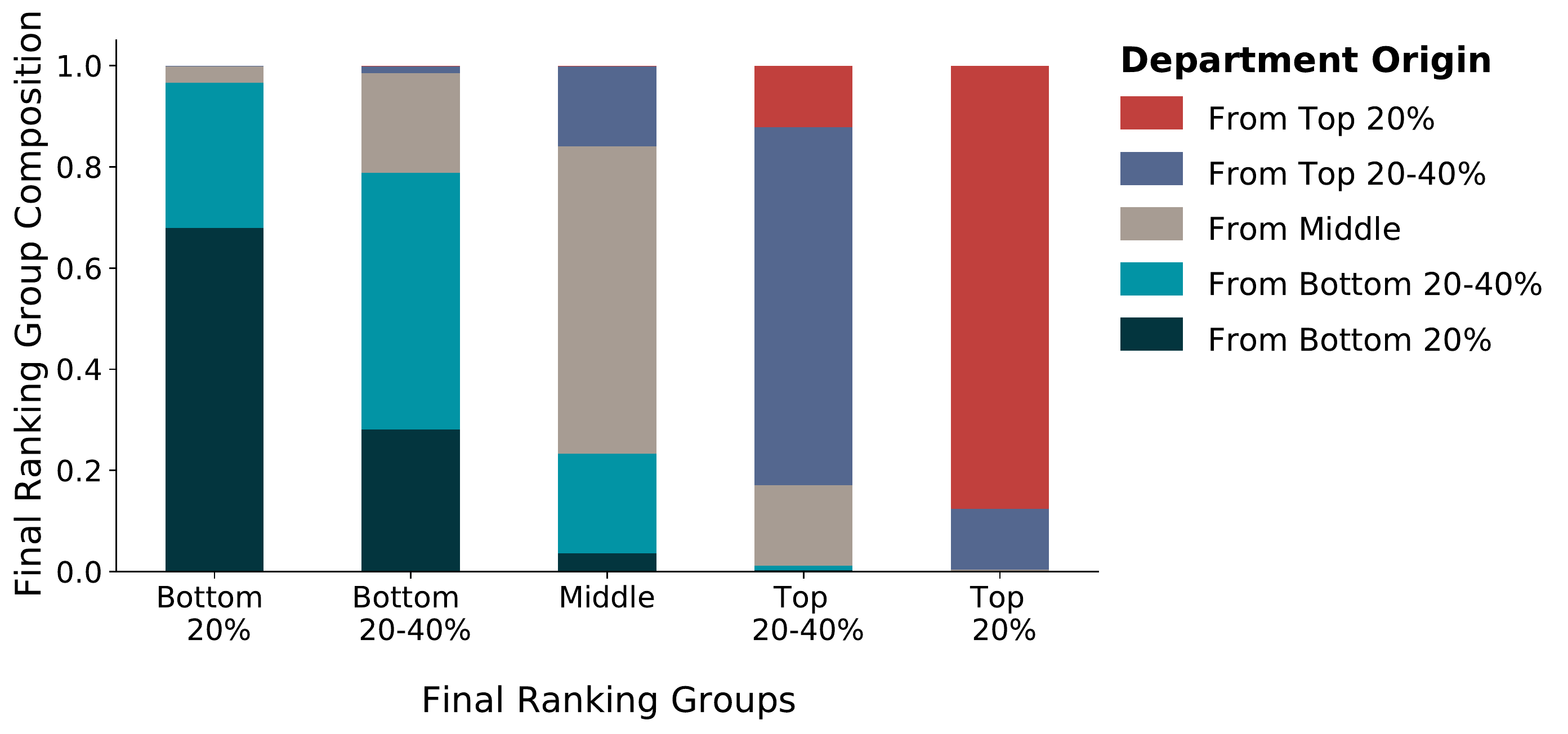}
\caption{\textbf{Final ranking group composition of Computer Science departments.} Rank mobility results derived from the best fitting Faculty production model with egalitarian initialization, with $\beta = 1.0$ and $p = 0.02$. The composition of departments' origin is traced and averaged over $50$ hiring simulated networks.}
\label{Sfig:cs_robust}
\end{figure*}

\begin{figure*}[ht!]
\centering
\includegraphics[width=0.85\linewidth]{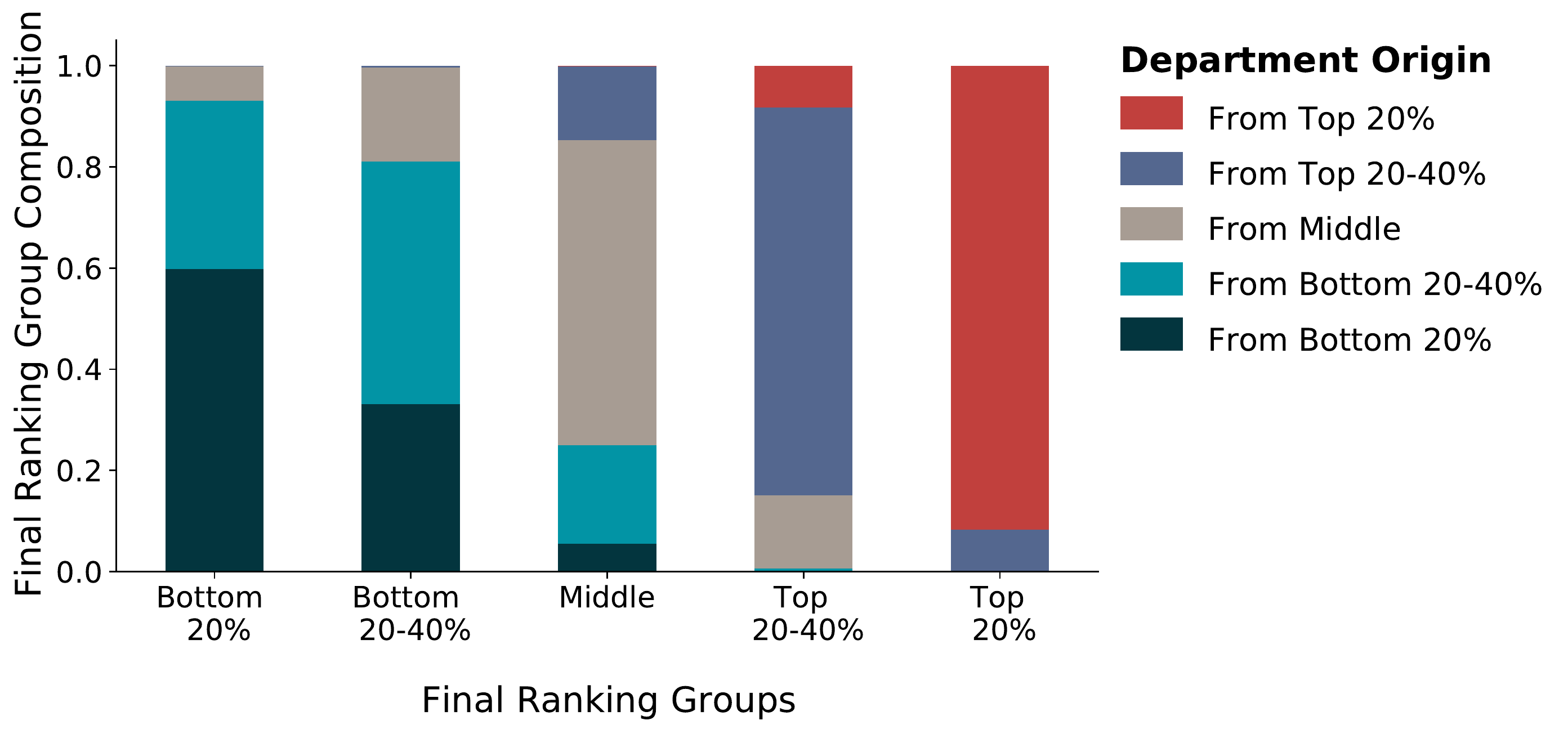}
\caption{\textbf{Final ranking group composition of History departments.} Rank mobility results derived from the best fitting Faculty production model with egalitarian initialization, with $\beta = 1.0$ and $p = 0.009$. The composition of departments' origin is traced and averaged over $50$ hiring simulated networks.}
\label{Sfig:hs_robust}
\end{figure*}

\end{document}